\documentclass[twoside,preprintnumbers,superscriptaddress,amsmath,amssymb,11pt,pacs,nofootinbib]{revtex4}
\usepackage{amsmath,amssymb,url}
\usepackage{graphicx,pslatex,subfigure}
\usepackage{esint}
\usepackage{lscape}
\usepackage{pstricks}
\usepackage{color}
\usepackage{braket,euscript}
\usepackage{fancyvrb}

\newcommand{\p}{\partial}

\begin{document}

\title{{\bf  A local and BRST-invariant Yang-Mills theory within the Gribov horizon}}
\author{M.~A.~L.~Capri}
\email{caprimarcio@gmail.com}
\affiliation{Departamento de F\'{\i}sica Te\'{o}rica,
Instituto de F\'{\i}sica, UERJ - Universidade do Estado do Rio de Janeiro, Rua S\~ao Francisco Xavier 524, 20550-013, Maracan\~a, Rio de Janeiro, Brasil}\author{D.~Dudal}
\email{david.dudal@kuleuven.be}
\affiliation{KU Leuven Campus Kortrijk - KULAK, Department of Physics, Etienne Sabbelaan 51 bus 7800, 8500 Kortrijk, Belgium}
\affiliation{Ghent University, Department of Physics and Astronomy, Krijgslaan 281-S9, 9000 Gent, Belgium}
\author{D.~Fiorentini}
\email{diegofiorentinia@gmail.com}
\affiliation{Departamento de F\'{\i}sica Te\'{o}rica,
Instituto de F\'{\i}sica, UERJ - Universidade do Estado do Rio de Janeiro, Rua S\~ao Francisco Xavier 524, 20550-013, Maracan\~a, Rio de Janeiro, Brasil}\author{M.~S.~Guimaraes}
\email{msguimaraes@uerj.br}
\affiliation{Departamento de F\'{\i}sica Te\'{o}rica,
Instituto de F\'{\i}sica, UERJ - Universidade do Estado do Rio de Janeiro, Rua S\~ao Francisco Xavier 524, 20550-013, Maracan\~a, Rio de Janeiro, Brasil}\author{I.~F.~Justo}
\email{igorfjusto@gmail.com}
\affiliation{Departamento de F\'{\i}sica Te\'{o}rica,
Instituto de F\'{\i}sica, UERJ - Universidade do Estado do Rio de Janeiro, Rua S\~ao Francisco Xavier 524, 20550-013, Maracan\~a, Rio de Janeiro, Brasil}\affiliation{Ghent University, Department of Physics and Astronomy, Krijgslaan 281-S9, 9000 Gent, Belgium}\affiliation{UFES - Universidade Federal do Esp\'irito Santo, Departamento de Qu\'imica e F\'isica, Centro de Ci\^encias Exatas, Naturais e da Sa\'ude, Alto Universit\'ario, s/n,29500-000, Gurarema, Alegre, ES, Brasil.}
\author{A.~D.~Pereira}
\email{aduarte@if.uff.br}
\affiliation{UFF $-$ Universidade Federal Fluminense, Instituto de F\'{\i}sica, Campus da Praia Vermelha, Avenida General Milton Tavares de Souza s/n, 24210-346, Niter\'oi, RJ, Brasil.}
\affiliation{Max Planck Institute for Gravitational Physics (Albert Einstein Institute), Am M\"uhlenberg 1, Potsdam, 14476 Germany}
\author{B.~W.~Mintz}
\email{bruno.mintz.uerj@gmail.com}
\affiliation{Departamento de F\'{\i}sica Te\'{o}rica,
Instituto de F\'{\i}sica, UERJ - Universidade do Estado do Rio de Janeiro, Rua S\~ao Francisco Xavier 524, 20550-013, Maracan\~a, Rio de Janeiro, Brasil}\author{L.~F.~Palhares}
\email{leticia.palhares@uerj.br}
\affiliation{Departamento de F\'{\i}sica Te\'{o}rica,
Instituto de F\'{\i}sica, UERJ - Universidade do Estado do Rio de Janeiro, Rua S\~ao Francisco Xavier 524, 20550-013, Maracan\~a, Rio de Janeiro, Brasil}\author{R.~F.~Sobreiro}
\email{sobreiro@if.uff.br}
\affiliation{UFF $-$ Universidade Federal Fluminense, Instituto de F\'{\i}sica, Campus da Praia Vermelha, Avenida General Milton Tavares de Souza s/n, 24210-346, Niter\'oi, RJ, Brasil.}
\author{S.~P.~Sorella}
\email{silvio.sorella@gmail.com}
\affiliation{Departamento de F\'{\i}sica Te\'{o}rica,
Instituto de F\'{\i}sica, UERJ - Universidade do Estado do Rio de Janeiro, Rua S\~ao Francisco Xavier 524, 20550-013, Maracan\~a, Rio de Janeiro, Brasil}

%\date{\today}
\begin{abstract}
We present a  local setup for the recently introduced BRST-invariant formulation of Yang-Mills theories for  linear covariant gauges that takes into account the existence of gauge copies {\it \`a la} Gribov and Zwanziger. Through the convenient use of auxiliary fields, including one of the Stueckelberg  type, it is shown that both the action and the associated nilpotent BRST operator can be put in local form. Direct consequences of this fully local and BRST-symmetric framework are drawn from its Ward identities: (i)
an exact prediction for the longitudinal part of the gluon propagator in  linear covariant gauges that is compatible with recent lattice results and (ii) a proof of the gauge-parameter independence of all correlation functions of  local BRST-invariant operators.
\end{abstract}
\maketitle

%%%%%%%%%%%%%%	

\section{Introduction}

Yang-Mills  gauge  theory and its extensions including matter fields are extremely successful perturbatively \cite{Weinberg:1996kr}.  One of the formal grounds  for this success is a path-integral formulation based on the Faddeev-Popov method to perturbatively fix the gauge and allow for Feynman diagram computations.  There is an inherent Becchi-Rouet-Stora-Tyutin (BRST) symmetry.

In Landau gauge, $\p_\mu A_\mu^a=0$,  and in Euclidean spacetime, the gauge path integral takes the form \footnote{We adopt standard notation, which can be found e.g.~in Ref. \cite{Weinberg:1996kr}.}:
\begin{eqnarray}
	&&\mathcal{Z}=\int \left[\EuScript{D}\mathbf{A}\right]\; \delta(\partial A^a) \; \det({\cal M})\;\mathrm{e}^{- S_{\mathrm{YM}} }
	=\int \left[\EuScript{D}\mathbf{A}\right]\; \mathrm{e}^{- S_{\mathrm{FP}} }
	\nonumber\\
&& S_{\mathrm{FP}}=S_{\mathrm{YM}}+ \int d^4x \left(i b^a \partial_\mu A^a_\mu + {\bar c}^a \partial_\mu D^{ab}_\mu c^b \right) \,,
\label{psetup}
\end{eqnarray}
and  it  is invariant under  the  BRST transformations:
\begin{eqnarray}
s A^a_\mu &=&  - D_\mu^{ab} c^b \;,~ s c^a = \frac{g}{2} f^{abc} c^b c^c   \;,~ s {\bar c}^a  =   i b^a  \;, ~s b^a = 0
 \;.
\end{eqnarray}
This perturbative setup allows for the establishment of several crucial features, ranging from all-order renormalizability \footnote{To be precise, this statement is actually restricted to the class of gauge fixings for which the quantum action principle \cite{Piguet:1995er} has been fully established. An example of such class of gauges is the one which respects Lorentz covariance. For noncovariant gauges, the situation can be much more involved \cite{Gaigg:1990si,Bassetto:1991ue,Leibbrandt:1994wj}.} and gauge independence to perturbative unitarity, as well as a consistent extension for the physics of massive gauge bosons, through the Brout-Englert-Higgs mechanism.

Non-perturbatively, however, the situation is by far not the same. While numerical methods provide strong evidence for confinement and dynamical chiral symmetry breaking (when fermions are included) in the strong-coupling region \cite{Rothe:1992nt,Durr:2008zz}, this regime of the theory is not amenable to a fully analytical scrutiny, leaving many theoretical gaps concerning e.g.~the  possible  mechanisms that can  drive such remarkable physical phenomena.

An important drawback of the Faddeev-Popov setup in the nonperturbative regime was first pointed out by Gribov \cite{Gribov:1977wm}: the sampling over physically inequivalent gauge configurations in the path integral is hindered by the presence of gauge copies in the gauge fixing procedure
(for recent reviews, cf. \cite{Sobreiro:2005ec,Vandersickel:2012tz}). Away from the perturbative vicinity in gauge configuration space, large fields and eventually zero modes of the Faddeev-Popov operator,
	\begin{equation}
	\mathcal{M}^{ab}=-\delta^{ab}\partial^2+gf^{abc}A^{c}_{\mu}\partial_{\mu},\,\,\,\, \mathrm{with}\,\,\,\, \partial_{\mu}A^{a}_{\mu}=0 \textrm{   (Landau gauge)}\,,
	\label{intro0}
	\end{equation}
are reached and the functional integral measure
in \eqref{psetup} becomes ill-defined,  which can be directly appreciated from the presence of $\det(\mathcal{M})=0$.
Over the last decades advances have been made in the direction of circumventing this problem. In particular, the Gribov-Zwanziger framework \cite{Zwanziger:1989mf}  encodes a modification of the Faddeev-Popov procedure in order to account for the presence of gauge copies. It
corresponds to restricting the functional measure to the first Gribov region
$\Omega$ in the Landau gauge, $\partial_\mu A^a_\mu=0$, namely
\begin{equation}
\Omega = \{ \; A^a_\mu|  \;  \partial_\mu A^a_\mu =0,  \;    {\mathcal M}^{ab}(A) > 0 \; \}    \;, \label{om}
\end{equation}
where ${\mathcal{M}}^{ab}$ is the Faddeev-Popov operator \eqref{intro0}.
In practice, a nonlocal horizon term is introduced in the SU($N$) gauge action using a new mass parameter $\gamma$ -- the Gribov parameter:
	\begin{eqnarray}
S_{\mathrm{FP}} \mapsto S_{\mathrm{GZ}}=S_{\mathrm{FP}} +\gamma^4H(A) - 4V\gamma^4(N^2-1)
	\label{gz1}
	\end{eqnarray}
	where $V$ is the spacetime volume and the horizon function reads
	\begin{equation}
	H(A)=g^2\int d^4xd^4y~f^{abc}A^{b}_{\mu}(x)\left[\mathcal{M}^{-1}(x,y)\right]^{ad}f^{dec}A^{e}_{\mu}(y)\,,
	\label{intro3}
	\end{equation}
	in terms of the inverse of the Faddeev-Popov operator, $\left[\mathcal{M}^{-1}\right]$.
The constraint that  enables us to get rid of the gauge copies\footnote{We remind here that the Gribov region $\Omega$ is not free from Gribov copies, {\it i.e.} additional copies still exist inside $\Omega$.   A smaller region within $\Omega$ exists which is fully free from Gribov copies. This region is called  the Fundamental Modular Region. Though, unlike the case of  the Gribov region $\Omega$, a local and renormalizable framework implementing the restriction to the Fundamental Modular Region is, at present,  unknown. Therefore, we shall proceed by focusing on the region $\Omega$. } is then realized by fixing the Gribov parameter $\gamma$ in a self-consistent way through the extremization of the vacuum energy of the theory with respect to $\gamma^2$, yielding the gap equation:
	\begin{equation}
	\langle H \rangle = 4V(N^2-1)\,,
	\label{intro4}
	\end{equation}
where the vacuum expectation value $\langle H \rangle$ has to be evaluated with the measure defined by the action in eq.~\eqref{gz1}.

Even though this approach and its subsequent extensions have provided interesting results for correlation functions of fundamental fields \cite{Dudal:2007cw,Dudal:2008sp,Dudal:2011gd,Baulieu:2009xr}
as well as for physical observables like the bound state spectrum  \cite{Dudal:2010cd,Dudal:2013wja,Dudal:2013vha}, the thermodynamics  and transport properties of the quark-gluon plasma  \cite{Canfora:2015yia,Canfora:2013kma,Canfora:2013zna,Lichtenegger:2008mh,Fukushima:2012qa,Fukushima:2013xsa,Su:2014rma,Florkowski:2015dmm,Bandyopadhyay:2015wua,Guimaraes:2015vra},
the standard BRST  symmetry  is  softly broken. The absence of this nilpotent symmetry that plays a crucial role in the perturbative
setup  can obscure  features such as renormalizability, gauge independence, definition of a physical space and unitarity in the infrared regime of the theory. It has been established, however, that a softly broken BRST operator is sufficient in several examples to guarantee that a gauge theory is renormalizable,  thus still displaying  predictive power. The other issues are still under intense investigation. For recent and different perspectives on this important topic, the reader is referred to e.g.~\cite{Dudal:2009xh,Sorella:2009vt,Baulieu:2008fy,Capri:2010hb,Dudal:2012sb,Dudal:2014rxa,Capri:2012wx,Pereira:2013aza,Pereira:2014apa,Capri:2014bsa,Tissier:2010ts,Serreau:2012cg,Serreau:2015yna,Lavrov:2013boa,Moshin:2015gsa,Schaden:2014bea,Cucchieri:2014via,Cucchieri:2016czg,Dudal:2010hj}.

Here, we concentrate on a recent development put forward in Refs.~\cite{Capri:2015ixa,Capri:2015nzw}: the existence of a nonperturbative BRST operator that is both nilpotent and an exact  symmetry of the Gribov-Zwanziger action in Landau and linear covariant gauges. The proposal relies on rewriting the Landau-gauge horizon function, \eqref{intro3}, in terms of a non-local gauge-invariant transverse field $A_{\mu }^{h}$, namely
\begin{equation}
H(A) = H(A^h) - R(A) (\partial A) = H(A^h) - \int d^4x d^4y R^a(x,y) (\partial A^a)_y
\end{equation}
where
	\begin{eqnarray}
	A_{\mu }^{h} &=&P_{\mu\nu} \left( A_{\nu }-ig\left[ \frac{\partial A}{\partial
		^{2}},A_{\nu }\right] +\frac{ig}{2}\left[ \frac{\partial A}{\partial ^{2}}%
	,\partial _{\nu }\frac{\partial A}{\partial ^{2}}\right] \right)
	+O(A^{3})  \nonumber \\
	&=&A_{\mu }-\frac{\partial _{\mu }}{\partial ^{2}}\partial A+ig\left[ A_{\mu
	},\frac{1}{\partial ^{2}}\partial A\right] +\frac{ig}{2}\left[ \frac{1}{%
	\partial ^{2}}\partial A,\partial _{\mu }\frac{1}{\partial ^{2}}\partial
A\right]\nonumber\\&& +ig\frac{\partial _{\mu }}{\partial ^{2}}\left[ \frac{\partial
	_{\nu }}{\partial ^{2}}\partial A,A_{\nu }\right]   +i\frac{g}{2}\frac{\partial _{\mu }}{\partial ^{2}}\left[ \frac{\partial A%
}{\partial ^{2}},\partial A\right] +O(A^{3})\,,  \label{hhh3g}
\end{eqnarray}
with $P_{\mu\nu}=\left(\delta _{\mu \nu }-\frac{\partial _{\mu }\partial_{\nu }}{\partial ^{2}}\right) $ being the transverse projector. By construction, expression \eqref{hhh3g} is left invariant by gauge transformations order by order.  This field $A_\mu^h$ can be obtained through the minimization of $\int d^4x A_\mu^a A_\mu^a$ along its gauge orbit. We refer to \cite{Capri:2015ixa} and references therein for details.

The term $R(A)(\partial A)$ is an infinite non-local power series of $A_\mu$ that collects all terms proportional to  $(\partial_{\mu} A_{\mu}^a)$ and can be reabsorbed by a shift of the auxiliary field $b$ as follows
\begin{eqnarray}
 S_{GZ} &=& S_{YM} + \int d^4x \left(  i b^a \partial_\mu A^a_\mu + {\bar c}^a \partial_\mu D^{ab}_\mu c^b \right)  + \gamma^4 H(A)
    \nonumber \\
& = & S_{YM} + \int d^4x \left( \ i b^{h,a} \partial_\mu A^a_\mu + {\bar c}^a \partial_\mu D^{ab}_\mu c^b \right)  + \gamma^4 H(A^h)      \;,
\label{gzh2}
\end{eqnarray}
where the new field $b^h$ reads
\begin{equation}
b^h = b + i  \gamma^4 R(A)   \;. \label{bh}
\end{equation}
The action \eqref{gzh2}, with the horizon function $H(A^h)$ expressed via Zwanziger's auxiliary fields  $({\bar \varphi}, \varphi, \omega, {\bar \omega})$, i.e.
\begin{eqnarray}
\gamma^4 H(A^h)
\mapsto
  \int d^4x\;  \left( - {\bar \varphi} {{\cal M}}(A^h) \varphi + {\bar \omega} {{\cal M}}(A^h) \omega + \gamma^2 A^h ({\bar \varphi} + \varphi)   \right), \label{gzh3}
\end{eqnarray}
enjoys the following exact nilpotent BRST symmetry:
\begin{equation}
s_{\gamma^2} =  s + \delta_{\gamma^2}\,,\qquad  s_{\gamma^2}^2=0\;, \qquad
s_{\gamma^2} S_{GZ} = 0 \;. \label{gzinv}
\end{equation}
where the operator $s$ stands for  the usual BRST operator
\begin{eqnarray}
s A^a_\mu &=&  - D_\mu^{ab} c^b \;,~ s c^a = \frac{g}{2} f^{abc} c^b c^c   \;,~ s {\bar c}^a  =  i  b^a  \;, ~s b^a = 0      \;, \nonumber \\
s \varphi_\mu^{ab} & = & \omega_\mu^{ab} \;, ~ s \omega_\mu^{ab}=0 \;, ~ s {\bar \omega}_\mu^{ab} =   {\bar \varphi}_\mu^{ab}   \;, ~ s{\bar \varphi}_\mu^{ab} =0   \;, \label{brst3}
\end{eqnarray}
while
\begin{eqnarray}
\delta_{\gamma^2} {\bar c}^a & = & - \gamma^4 R^a(A)  \;,\quad \delta_{\gamma^2}b^{h,a} = - i \gamma^4 sR^a(A) \;,\nonumber\\
\delta_{\gamma^2} {\bar \omega}_\mu^{ac}& =&  -  \gamma^2 gf^{kbc} A_\mu^{h,k}\left[{\cal M}^{-1}(A^h)\right]^{ba}\;,~\delta_{\gamma^2}(\text{rest})=0\;.
\label{deltadef}
\end{eqnarray}
The presence of the Gribov parameter in the definition of this new exact BRST transformation makes explicit its nonperturbative nature, while guarantees that the perturbative limit is fully recovered in the ultraviolet, as desired. Furthermore, the extension of this framework to the class of  linear covariant gauges becomes a straightforward procedure based on the BRST principle  \cite{Baulieu:1981sb}, in complete analogy with what is done in the perturbative case. The resulting action is
\cite{Capri:2015ixa}:
\begin{equation}
 S_{GZ}^{LCG} = S_{FP}^{h} + \int d^4x\;  \left( - {\bar \varphi}^{ac}_{\mu} {{\cal M}}^{ab}(A^h) \varphi^{bc}_{\mu} +  {\bar \omega}^{ac}_{\mu} {{\cal M}}^{ab}(A^h) \omega^{bc}_{\mu} +g \gamma^2 f^{abc}A^{h,a}_{\mu} ({\bar \varphi} + \varphi)^{bc}_{\mu}   \right)\;, \label{fp88a}
\end{equation}
with
\begin{eqnarray}
S_{FP}^{h} &=& S_{YM} + s_{\gamma^2}\int d^4x \left( {\bar c}^{a} \p_\mu A^{a}_\mu -  i \frac{\alpha}{2} {\bar c}^ab^{h,a}\right)\nonumber\\       \label{fpLCG}
&=& S_{YM}+ \int d^4x \left(  i b^{h,a} \p_\mu A^a_\mu + \frac{\alpha}{2}b^{h,a} b^{h,a} +{\bar c}^{a} \p_\mu D^{ab}_\mu c^b\right)
\end{eqnarray}
 The parameter $\gamma^2$ is still to be determined from the horizon condition, see \eqref{intro3}-\eqref{intro4}, though with $A_\mu$ replaced by its gauge invariant counterpart $A_\mu^{h}$. As such, the horizon condition itself becomes a gauge-invariant condition.

In this fashion, as discussed in \cite{Capri:2015ixa}, a substantial set of (infinitesimally) gauge equivalent configurations, obeying the same linear covariant gauge condition, are excluded from the path integration. More precisely, all infinitesimal gauge copies that have a Taylor expansion around the Landau gauge, $\alpha=0$, are eliminated.

Finally, we end up with a nonperturbative formulation of the Gribov-Zwanziger theory for  linear covariant gauges that displays an exact, nilpotent BRST invariance. Nevertheless, the full power of this symmetry is established only in local quantum field theories, which is not the case of this one so far. Indeed, inspection of eqs.\eqref{gzinv}--\eqref{fp88a} reveals two sources of non-locality in the action and in the definition of the new BRST transformation: the gauge-invariant field $A^h$ and the inverse of the Faddeev-Popov operator.

In this work, we show that this novel perspective on Yang-Mills  theories in the presence of the Gribov horizon can be cast in a fully local form -- with both the action and its exact nonperturbative BRST symmetry localizable. The first consequences of this fully local version will also be explored, shedding some light on the most relevant issue of gauge independence as well as providing an exact, all-order result for the longitudinal part of the gluon propagator in linear covariant gauges.

This paper is organized as follows. In the next section, the action \eqref{fp88a} for the linear covariant gauges is shown to be localized via the introduction of an auxiliary field of the Stueckelberg type. Section \ref{section:LocalBRST} then presents the procedure for obtaining a local form for the nonperturbative BRST symmetry.
In Section \ref{section:Ward}, we write down Ward identities from which a set of interesting
consequences will arise, namely:
the gauge-parameter independence of BRST-invariant correlation functions and
an exact result for the longitudinal part of the gluon propagator, that turns out to be the same as
in perturbative Yang-Mills theory.
The extension of the framework that takes into account the formation of dimension-two condensates
is discussed in Section \ref{section:RGZ}, while the treatment of matter fields is addressed in Section
\ref{section:matter}.
Section \ref{section:Finaldiscussion} collects our conclusions and outlook.

%%%%%%%%%%%%%%%%%
\section{Localization of the BRST-invariant action}\label{loc-action}

An important step towards the construction of the theory is that of finding expressions of the action, its
fundamental fields, and the corresponding nilpotent BRST symmetry in terms of local fields only.
In this section, we write down the action as a local field theory, while the local version of its BRST
symmetry will be discussed in the next section.

We first notice that the horizon function $H(A^h)$ in the action (\ref{gzh2}) has a double source of
non-locality. First, the horizon function is itself a non-local functional of its argument due to the
presence of the inverse of the Fadeev-Popov operator $({\cal M}^{-1})^{ab}$. Second, the field $A^h$
has so far been formulated as a nonlocal series in the gauge field $A$, eq.\eqref{hhh3g}. As discussed in the previous section, the non-local quantity  $H(A^h)$ can be re-expressed by introducing  bosonic fields
$(\bar\varphi^{ab}_\mu,\varphi^{ab}_\mu)$, as well as fermionic fields $(\bar\omega^{ab}_\mu,\omega^{ab}_\mu)$.
The corresponding action reads \footnote{We point out that the Lagrange multiplier  $b^h$ appearing in expression \eqref{fpLCG} can be considered as an elementary field. This follows by noticing that the field redefinition in eq. \eqref{bh}  corresponds in fact to a field transformation with unit Jacobian. From now on we shall remove the index $h$ in the Lagrange multiplier field $b$.}
\begin{eqnarray}
 S_{GZ}^{LCG} &=& S_{YM} + \int d^4x \left( \alpha\frac{b^ab^a}{2} + ib^a \partial_\mu A_\mu^a + {\bar c}^a \partial_\mu D^{ab}_\mu(A) c^b \right)
    \;\nonumber\\
&& \hspace{1cm} + \int\,d^4x \left(-\bar\varphi^{ac}_\mu{\cal M}^{ab}(A^h)\varphi^{bc}_\mu + \bar\omega^{ac}_\mu{\cal M}^{ab}(A^h)\omega^{bc}_\mu
+ g\gamma^2\,f^{abc}(A^h)^a_\mu(\varphi_\mu^{bc} + \bar\varphi^{bc}_\mu)\right),
\label{gzlocalhorizon}
\end{eqnarray}
in which the nonlocality stemming from the integral operator ${\cal M}^{-1}$ present in the horizon function
has been traded  for  the new auxiliary fields $(\bar\varphi^{ab}_\mu,\varphi^{ab}_\mu)$ and
$(\bar\omega^{ab}_\mu,\omega^{ab}_\mu)$.

Our next step towards the definition of a BRST-invariant Gribov-Zwanziger theory is to write its
action in terms of local fields only. As introduced in \cite{Capri:2015ixa} and briefly reviewed in
the Introduction, the field $A^h$ has been originally defined as a highly nonlocal series in the gauge field.
In order to write a local representation for the gauge-invariant field $A^h$ and consequently find a local action,
%\footnote{We employ here already the shifted field $b^h=b - R(A)$, {\it i.e.} $b\rightarrow b^h$, whose Jacobian is unity.}
we shall employ a different representation of it, using an auxiliary Stueckelberg field $\xi^a$, see refs.\cite{Delbourgo:1986wz,Delbourgo:1987np,Lavelle:1995ty,Dragon:1996tk,Ruegg:2003ps}.
The field $\xi^a$ can be introduced by first defining
\begin{equation}
h=e^{ig\,\xi^{a}T^{a}}  \;,
\label{hxi}
\end{equation}
so that we write the field $A^h$ as
\begin{equation}
 A^{h}_{\mu}=(A^{h})^{a}_{\mu}T^{a}=h^{\dagger}A^{a}_{\mu}T^{a}h+\frac{i}{g}\,h^{\dagger}\partial_{\mu}h\,.    \label{st}
\end{equation}

An important feature of $A^h$, as defined by \eqref{st}, is that it is gauge invariant, that is,
\begin{equation}
A^{h}_{\mu} \rightarrow A^{h}_{\mu} \;,
\end{equation}
as can be seen from the gauge transformations  with SU($N$) matrix $V$
\begin{equation}
A_\mu \rightarrow V^{\dagger} A_\mu V + \frac{i}{g} V^{\dagger} \partial_ \mu V
\;, \qquad h \rightarrow V^{\dagger} h     \;, \qquad h^{\dagger} \rightarrow h^{\dagger} V \;.
\end{equation}

The field $A^{h}_\mu$ is now a local field and can be expanded in terms of $\xi^a$, yielding
\begin{equation}
(A^{h})^{a}_{\mu}=A^{a}_{\mu}-D^{ab}_{\mu}\xi^{b}-\frac{g}{2}f^{abc}\xi^{b}D^{cd}_{\mu}\xi^{d}+\mathcal{O}(\xi^{3})\,.
\end{equation}
This construction allows us to write a local action for the Gribov-Zwanziger theory
in  linear covariant gauges,
\begin{eqnarray}\label{lact}
S_{GZ}^{LCG}  &=& \int d^{4}x  \,\bigg\{\, \frac{1}{4} \,F^{a}_{\mu\nu}F^{a}_{\mu\nu}+\frac{\alpha}{2}\,b^{a}b^{a}
+ib^{a}\,\partial_{\mu}A^{a}_{\mu}
+\bar{c}^{a}\partial_{\mu}D^{ab}_{\mu}(A)c^{b}  \nonumber \\
 &-& \bar\varphi^{ac}_{\nu}{\cal M}^{ab}(A^h)\varphi^{bc}_{\nu}
+\bar\omega^{ac}_{\nu}{\cal M}^{ab}(A^h)\omega^{bc}_{\nu}
+\gamma^{2}g\,f^{abc}(A^h)^{a}_{\mu}(\varphi^{bc}_{\mu}+\bar\varphi^{bc}_{\mu})+\tau^{a}\,\partial_{\mu}(A^h)^{a}_{\mu}\,\bigg\}\,.
\end{eqnarray}
Notice that the action is now written in terms of local fields only, with the Faddeev-Popov operator
${\cal M}^{ab}(A^h) = - \partial_\mu D_\mu^{ab}(A^h)$ now in terms of the local version of $A^h$,
eq. \eqref{st}. The Lagrange multipliers $b^{a}$ and $\tau^{a}$ are needed to enforce two constraints.
The first one is the linear covariant gauge condition $\partial_{\mu}A^{a}_{\mu}=i\alpha\,b^{a}\,$,
while the second is the transversality of the field $A^h$, $\partial_{\mu}(A^h)^{a}_{\mu}=0\,$, which
can be seen as a  constraint  on the Stueckelberg field. Indeed, if the Stueckelberg field $\xi^a$ is
eliminated through the transversality
constraint $\partial_{\mu}(A^{h})^{a}_{\mu}=0$, we get back the non-local
expression for the field $A^h_\mu$ appearing in the action \eqref{gzlocalhorizon}.
By further integrating  over  the auxiliary fields
$(\bar\varphi^{ab}_\mu,\varphi^{ab}_\mu)$ and $(\bar\omega^{ab}_\mu,\omega^{ab}_\mu)$,
one goes back to the original action (\ref{gzh2}).

Expression \eqref{lact}  is now local, albeit non-polynomial.
To some extent, the situation shares similarities with other local non-polynomial
models like ${\cal N}=1$ SYM in superspace, non-linear two-dimensional sigma models as well
as chiral Wess-Zumino models  \cite{Piguet:1984js,Gates:1983nr,Wess:1973kz,Rajaraman:1982is}). In all these cases the algebraic renormalization proved
to be very helpful, allowing in fact for an all order proof of the renormalizability
of these models. Perhaps, this will be also the case of our model, a thought reinforced
by the non-trivial UV finiteness properties displayed by the horizon function and by
Gribov-like propagators \cite{Capri:2015mna}. Let us also remind that the renormalizability of Landau gauge Yang-Mills theories supplemented by a Stueckelberg mass term has been proven in \cite{Delbourgo:1986wz,Delbourgo:1987np}.

%%%%%%%%%%%%%%%%%%%%%%%%%%%

\section{Local BRST transformations}\label{section:LocalBRST}

Now that we have a local version of the  nonperturbative BRST-invariant GZ action, in the next  subsections we
 address the localization of its nonperturbative BRST transformations.

%%%%
\subsection{The nonperturbative BRST operator $s_{\gamma^2}$}

For the nilpotent BRST transformations we have
\begin{eqnarray}
sA^{a}_{\mu}&=&-D^{ab}_{\mu}c^{b}\,,\\
sc^{a}&=&\frac{g}{2}f^{abc}c^{b}c^{c}\,,\\
s\bar{c}^{a}&=&ib^{a}\,,\\
sb^{a}&=&0\,.   \label{brst}
\end{eqnarray}
Following \cite{Dragon:1996tk}, for the Stueckelberg field we write,  with $i,j$ indices associated with a generic representation,
% the fundamental representation,
%
\begin{equation}
s h^{ij} = -ig c^a (T^a)^{ik} h^{kj}  \;, \qquad s (A^h)^a_\mu = 0  \;,  \label{brstst}
\end{equation}
from which the BRST transformation of the field $\xi^a$ (cf. eq. \eqref{hxi}) can be evaluated iteratively, giving
\begin{equation}
s \xi^a=  - c^a + \frac{g}{2} f^{abc}c^b \xi^c - \frac{g^2}{12} f^{amr} f^{mpq} c^p \xi^q \xi^r + O(g^3)    \;.
\label{eqsxi}
\end{equation}
It is instructive to check here explicitly the BRST invariance of $A^h$. For this, it is better to employ a
matrix notation for the fields, namely
\begin{eqnarray}
sA_\mu &=& -\partial_\mu c + ig [A_\mu, c] \;, \qquad s c = -ig c c     \;, \nonumber \\
s h & =& -igch \;, \qquad sh^{\dagger}  = ig h^{\dagger} c  \;, \label{mbrst}
\end{eqnarray}
with $A_\mu = A^a_\mu T^a$, $c=c^a T^a$, $\xi=\xi^a T^a$. From expression  \eqref{st} we get
\begin{eqnarray}
s A^h_\mu & = & ig h^{\dagger} c \;A_\mu h + h^{\dagger}  (-\partial_\mu c + ig [A_\mu, c]) h -ig h^{\dagger} A_\mu \;c h - h^{\dagger} c \partial_\mu h + h^{\dagger} \partial_\mu(c h) \nonumber \\
&=& igh^{\dagger} c A_\mu h -  h^{\dagger}  (\partial_\mu c ) h +ig  h^{\dagger} A_\mu \;c h - ig  h^{\dagger} c \;A_\mu h -ig h^{\dagger} A_\mu c h - h^{\dagger} c \partial_\mu h + h^{\dagger} (\partial_\mu c) h + h^{\dagger} c \partial_\mu h   \nonumber \\
&=& 0 \;. \label{sah}
\end{eqnarray}
For the other fields, we write
\begin{eqnarray}
s\varphi^{ab}_{\mu}&=&\omega^{ab}_{\mu}\,\Longrightarrow\, s\omega^{ab}_{\mu}=0\,,\\
s\bar\omega^{ab}_{\mu}&=&\bar\varphi^{ab}_{\mu}\,\Longrightarrow s\bar\varphi^{ab}_{\mu}=0\,,\\
s\tau^{a}&=&0.
\end{eqnarray}
Within the BRST framework just introduced, the action \eqref{lact} can be recast in the form
\begin{eqnarray}
S_{GZ}^{LCG}&=&\frac{1}{4}\int d^{4}x\,F^{a}_{\mu\nu}F^{a}_{\mu\nu}+s\int d^{4}x\,\left(
-\frac{i\alpha}{2}\,\bar{c}^{a}b^{a}+\bar{c}^{a}\,\partial_{\mu}A^{a}_{\mu}
-\bar\omega^{ac}_{\nu}{\cal M}^{ab}(A^h)\varphi^{bc}_{\nu}\right)\nonumber\\
\;\;\;&&\;\;\;+\;\;\;\gamma^{2}\int d^{4}x\, gf^{abc}(A^h)^{a}_{\mu}(\varphi^{bc}_{\mu}+\bar\varphi^{bc}_{\mu}) +\int d^4x \;\tau^{a}\,\partial_{\mu}(A^h)^{a}_{\mu}.
\label{act2}
\end{eqnarray}
One of the most interesting features of the action \eqref{act2} is that it enjoys a nonperturbative nilpotent BRST symmetry
\begin{equation}
s_{\gamma^2} S_{GZ}^{LCG} = 0 \;,\;\;\; s_{\gamma^2}^2 =0 \label{ex}
\end{equation}
with the following full set of transformations defined as
\begin{eqnarray}
s_{\gamma^2}A^{a}_{\mu}&=&-D^{ab}_{\mu}c^{b}\,,\;\;\;\; s_{\gamma^2}c^{a}=\frac{g}{2}f^{abc}c^{b}c^{c}\,,\nonumber\\
s_{\gamma^2}\bar{c}^{a}&=&ib^{a}\,,\;\;\;\;
s_{\gamma^2}b^{a}= 0\,.  \nonumber\\
s_{\gamma^2} h^{ij} &=& -ig c^a (T^a)^{ik} h^{kj}  \;, \;\;\;\;  s_{\gamma^2} (A^h)^a_\mu = 0\,, \;\;\;\; \nonumber\\
s_{\gamma^2}\varphi^{ab}_{\mu}&=&\omega^{ab}_{\mu}\,,\;\;\;\; s_{\gamma^2}\omega^{ab}_{\mu}=0\,,\nonumber\\
s_{\gamma^2}\bar\omega^{ab}_{\mu}&=&\bar\varphi^{ab}_{\mu}   -  \gamma^2 gf^{kcb} (A^h)_\mu^{k}\left[{\cal M}^{-1}(A^h)\right]^{ca} \,,\;\;\;\; s_{\gamma^2}\bar\varphi^{ab}_{\mu}=0\,,\nonumber\\
s_{\gamma^2}\tau^{a}&=&0.
 \label{brstgamma}
\end{eqnarray}

\subsection{The localization of $s_{\gamma^2}$}

As mentioned before, the nonperturbative nilpotent BRST transformation \eqref{brstgamma}  defines  a symmetry of the action \eqref{act2}. However, these transformations are non-local and to have a proper well-defined quantum field theory we have to construct a fully local formulation. This can be done by introducing another set of auxiliary fields. We first note that
\begin{equation}
\label{horizon-loc}
\int  {\cal D}\varphi  {\cal D}\bar{\varphi}  {\cal D}\omega  {\cal D}\bar{\omega} \;  e^{-\int\,d^4x \left(-\bar\varphi^{ac}_\mu{\cal M}(A^h)^{ab}\varphi^{bc}_\mu + \bar\omega^{ac}_\mu{\cal M}(A^h)^{ab}\omega^{bc}_\mu
+ g\frac{\gamma^2}{\sqrt{2}}\,f^{abc}(A^h)^a_\mu(\varphi_\mu^{bc} + \bar\varphi^{bc}_\mu)\right)} \sim e^{-\frac{\gamma^4}{2}\;  H(A^h) }
 \end{equation}
where the symbol $\sim$ means ``up to a  prefactor''. We can also write this in terms of other fields
\begin{equation}
\label{horizon-loc-2}
\int  {\cal D}\beta  {\cal D}\bar{\beta}  {\cal D}\zeta  {\cal D}\bar{\zeta} \;  e^{-\int\,d^4x \left(-\bar\beta^{ac}_\mu{\cal M}(A^h)^{ab}\beta^{bc}_\mu + \bar\zeta^{ac}_\mu{\cal M}(A^h)^{ab}\zeta^{bc}_\mu
- g\frac{\gamma^2}{\sqrt{2}}\,f^{abc}(A^h)^a_\mu(\beta_\mu^{bc} + \bar\beta^{bc}_\mu)\right)} \sim e^{-\frac{\gamma^4}{2}\;  H(A^h) }
 \end{equation}
where $(\beta, \bar{\beta} )$ are complex bosonic fields and  $(\zeta, \bar{\zeta} )$ are fermionic fields. Multiplying these expressions we obtain a new local form for the action \eqref{gzh2}
\begin{eqnarray}
 S_{GZ}^{LCG} &=& S_{YM} + \int d^4x \left( \alpha\frac{b^ab^a}{2} + ib^a \partial_\mu A_\mu^a + {\bar c}^a \partial_\mu D^{ab}_\mu(A) c^b \right) +\int d^4x \;\tau^{a}\,\partial_{\mu}(A^h)^{a}_{\mu}.
    \;\nonumber\\
&& \hspace{1cm} + \int\,d^4x \left(-\bar\varphi^{ac}_\mu{\cal M}(A^h)^{ab}\varphi^{bc}_\mu + \bar\omega^{ac}_\mu{\cal M}(A^h)^{ab}\omega^{bc}_\mu
+ g\frac{\gamma^2}{\sqrt{2}}\,f^{abc}(A^h)^a_\mu(\varphi_\mu^{bc} + \bar\varphi^{bc}_\mu)\right),
\nonumber\\
&& \hspace{1cm} +\int\,d^4x \left(-\bar\beta^{ac}_\mu{\cal M}(A^h)^{ab}\beta^{bc}_\mu + \bar\zeta^{ac}_\mu{\cal M}(A^h)^{ab}\zeta^{bc}_\mu
- g\frac{\gamma^2}{\sqrt{2}}\,f^{abc}(A^h)^a_\mu(\beta_\mu^{bc} + \bar\beta^{bc}_\mu)\right)\,.
\label{gzlocalhorizon2}
\end{eqnarray}
It is clear from the construction that \eqref{gzh2}, \eqref{gzlocalhorizon}, and \eqref{gzlocalhorizon2} are all equivalent, sharing the same physical content. The advantage of introducing the extra set of auxiliary fields is that we can now define a local BRST symmetry for this action. It is straightforward to check that the following local nonperturbative BRST transformations are a symmetry of the action \eqref{gzlocalhorizon2}
\begin{eqnarray}
s_{loc}A^{a}_{\mu}&=&-D^{ab}_{\mu}c^{b}\,,\;\;\;\; s_{loc}c^{a}=\frac{g}{2}f^{abc}c^{b}c^{c}\,,\nonumber\\
s_{loc}\bar{c}^{a}&=&ib^{a}\,,\;\;\;\;
s_{loc}b^{a}= 0\,.  \nonumber\\
s_{loc} h^{ij} &=& -ig c^a (T^a)^{ik} h^{kj}  \;, \;\;\;\; s_{loc} (A^h)^a_\mu = 0\,, \;\;\;\; \nonumber\\
s_{loc}\varphi^{ab}_{\mu}&=&\omega^{ab}_{\mu}\,,\;\;\;\; s_{loc}\omega^{ab}_{\mu}=0\,,\nonumber\\
s_{loc}\bar\omega^{ab}_{\mu}&=&\bar\varphi^{ab}_{\mu} + \bar\beta_\mu^{ab}  \,,\;\;\;\; s_{loc}\bar\varphi^{ab}_{\mu}=0\,,\nonumber\\
s_{loc}\tau^{a}&=&0\,,\nonumber\\
s_{loc}\bar\beta_\mu^{ab} &=& 0\;\;\;\;
 s_{loc}\beta_\mu^{ab} = \omega_\mu^{ab} \nonumber\\
 s_{loc}\zeta_\mu^{ab} &=& 0 \;\;\;\; s_{loc}\bar\zeta_\mu^{ab} = 0\,,
 \label{brstgammalocal}
\end{eqnarray}
with
\begin{equation}
s_{loc} S_{GZ}^{LCG} = 0 \;. \label{symm}
\end{equation}
The fields $(\bar\zeta, \zeta)$ have $-1$ and $1$ ghost number, respectively, while the fields $(\bar\beta, \beta)$ have $0$ ghost number.   We point out that  the earlier non-local BRST operator $s_{\gamma^2}$, eqs.\eqref{brstgamma}, can be recovered from the local operator $s_{loc}$ upon elimination of the auxiliary field ${\bar \beta}$ through the equations of motion of $\beta$.

Note also that the original non-local action \eqref{gzh2} was written in terms of  $\gamma^4$ and the formulation of the action \eqref{gzlocalhorizon2} displays only $\gamma^2$. It follows that this observation leads to a natural discrete symmetry related to the invariance of the theory under $\gamma^2 \rightarrow -\gamma^2$. In terms of the fields, this amounts to the exchange  of both sets of auxiliary localizing fields, more precisely, the action is invariant under
\begin{eqnarray}
\varphi^{ab}_{\mu}&\rightarrow&-\beta^{ab}_{\mu}\,,\;\;\;\;
\bar\varphi^{ab}_{\mu}\rightarrow -\bar\beta^{ab}_{\mu}\,,\nonumber\\
\beta^{ab}_{\mu}&\rightarrow&-\varphi^{ab}_{\mu}\,,\;\;\;\;
\bar\beta^{ab}_{\mu} \rightarrow -\bar\varphi^{ab}_{\mu}\,,\nonumber\\
\omega^{ab}_{\mu}&\rightarrow&-\zeta^{ab}_{\mu}\,,\;\;\;\;
 \bar\omega^{ab}_{\mu}\rightarrow -\bar\zeta^{ab}_{\mu}\,,\nonumber\\
\zeta^{ab}_{\mu}&\rightarrow&-\omega^{ab}_{\mu}\,,\;\;\;\;
 \bar\zeta^{ab}_{\mu}\rightarrow -\bar\omega^{ab}_{\mu}\,.
 \label{discretesymm}
\end{eqnarray}
At this point it is worth to note that there is a natural set of field variables that we can identify as being BRST singlets, {\it i.e.} quantities which belong to the cohomology  of the  BRST operator, that is,  BRST-closed fields which are not  BRST-exact. We note that
\begin{eqnarray}
s_{loc}\left( \varphi^{ab}_{\mu} + \beta_\mu^{ab} \right) &=& 2\omega^{ab}_{\mu}\,, \nonumber\\
s_{loc}\left( \varphi^{ab}_{\mu} - \beta_\mu^{ab} \right) &=& 0\,,
 \label{coho}
\end{eqnarray}
which hints us  to  introduce the combinations
\begin{eqnarray}
\kappa^{ab}_{\mu} = \frac{1}{\sqrt{2}}\left( \varphi^{ab}_{\mu} + \beta_\mu^{ab} \right) \,, \nonumber\\
\lambda^{ab}_{\mu} = \frac{1}{\sqrt{2}}\left( \varphi^{ab}_{\mu} - \beta_\mu^{ab} \right) \,.
 \label{coho2}
\end{eqnarray}
 and their conjugate counterparts.  In terms of these fields, the action becomes \eqref{gzlocalhorizon2}
\begin{eqnarray}
 S_{GZ}^{LCG} &=& S_{YM} + \int d^4x \left( \alpha\frac{b^ab^a}{2} + ib^a \partial_\mu A_\mu^a + {\bar c}^a (\partial_\mu D_\mu)^{ab} c^b \right)  +\int d^4x \;\tau^{a}\,\partial_{\mu}(A^h)^{a}_{\mu}.
    \;\nonumber\\
&& \hspace{1cm} + \int\,d^4x \left(-\bar\kappa^{ac}_\mu{\cal M}(A^h)^{ab}\kappa^{bc}_\mu + \bar\omega^{ac}_\mu{\cal M}(A^h)^{ab}\omega^{bc}_\mu \right)
\nonumber\\
&& \hspace{1cm} +\int\,d^4x \left(-\bar\lambda^{ac}_\mu{\cal M}(A^h)^{ab}\lambda^{bc}_\mu + \bar\zeta^{ac}_\mu{\cal M}(A^h)^{ab}\zeta^{bc}_\mu
+ g\gamma^2\,f^{abc}(A^h)^a_\mu(\lambda_\mu^{bc} + \bar\lambda^{bc}_\mu)\right),\nonumber\\
&=& S_{YM} + s_{loc} \int d^4x \left( -i\frac{\alpha}{2}\bar c^a b^a + \bar c^a \partial_\mu A_\mu^a  -\frac{1}{\sqrt{2}} \bar\omega^{ac}_\mu{\cal M}(A^h)^{ab}\kappa^{bc}_\mu \right)  +\int d^4x \;\tau^{a}\,\partial_{\mu}(A^h)^{a}_{\mu}.
    \;\nonumber\\
&& \hspace{1cm} +\int\,d^4x \left(-\bar\lambda^{ac}_\mu{\cal M}(A^h)^{ab}\lambda^{bc}_\mu + \bar\zeta^{ac}_\mu{\cal M}(A^h)^{ab}\zeta^{bc}_\mu
+ g\gamma^2\,f^{abc}(A^h)^a_\mu(\lambda_\mu^{bc} + \bar\lambda^{bc}_\mu)\right),
\label{gzlocalhorizon3}
\end{eqnarray}
where the transformations \eqref{brstgammalocal} now read
\begin{eqnarray}
s_{loc}A^{a}_{\mu}&=&-D^{ab}_{\mu}c^{b}\,,\;\;\;\; s_{loc}c^{a}=\frac{g}{2}f^{abc}c^{b}c^{c}\,,\nonumber\\
s_{loc}\bar{c}^{a}&=&ib^{a}\,,\;\;\;\;
s_{loc}b^{a}= 0\,.  \nonumber\\
s_{loc} h^{ij} &=& -ig c^a (T^a)^{ik} h^{kj}  \;, \;\;\;\; s_{loc} (A^h)^a_\mu = 0\,, \;\;\;\; \nonumber\\
s_{loc}\kappa^{ab}_{\mu}&=&\sqrt{2} \omega^{ab}_{\mu}\,,\;\;\;\; s_{loc}\omega^{ab}_{\mu}=0\,,\nonumber\\
s_{loc}\bar\omega^{ab}_{\mu}&=&\sqrt{2} \bar\kappa^{ab}_{\mu} \,,\;\;\;\; s_{loc}\bar\kappa^{ab}_{\mu}=0\,,\nonumber\\
s_{loc}\tau^{a}&=&0\,,\nonumber\\
s_{loc}\bar\lambda_\mu^{ab} &=& 0\;\;\;\;
 s_{loc}\lambda_\mu^{ab} = 0\nonumber\\
 s_{loc}\zeta_\mu^{ab} &=& 0 \;\;\;\; s_{loc}\bar\zeta_\mu^{ab} = 0\,,
 \label{brstgammalocal2}
\end{eqnarray}
from which it becomes apparent that the fields  $\left((A^h)^a_\mu, \bar\lambda_\mu^{ab}, \lambda_\mu^{ab}, \bar\zeta_\mu^{ab}, \zeta_\mu^{ab}, \tau^a \right)$ define BRST singlets.  Note also that the term containing the Gribov parameter $\gamma$ belongs to the cohomology of the BRST operator $s_{loc}$, namely
\begin{equation}
 s_{loc}  \frac{\partial S_{GZ}^{LCG}}{\partial \gamma^2} = s_{loc} \int d^4x \left( g\,f^{abc}(A^h)^a_\mu(\lambda_\mu^{bc} + \bar\lambda^{bc}_\mu)  \right) =0 \;, \qquad \frac{\partial S_{GZ}^{LCG}}{\partial \gamma^2} \neq s_{loc} \Delta \;,
\end{equation}
for some local integrated field polynomial $\Delta$.  This important property  highlights  the fact that $\gamma$ is a physical parameter. On the other hand the gauge parameter $\alpha$ will not enter physical quantities, since it is associated to a BRST-exact term.  As such, it can also not contribute to the gap equation defining $\gamma$.  Henceforth, $\gamma$ is gauge independent and thus allowed to enter the expectation values of gauge invariant quantities, including the vacuum energy.

% % % % % % % % % % % % % % % % % % % % % % % % % % % % % % % % %

%%%%%%%%%%%%%%%%%%
\subsection{BRST-invariant infrared regularizing mass for the Stueckelberg field $\xi$}\label{subsection:StueckelbergMass}

The tree-level propagator of the Stueckelberg field $\xi$ can be derived from the quadratic part of the final action, eq. \eqref{gzlocalhorizon3}. The result is shown in Appendix \ref{AppProps} to be $\langle \xi \xi \rangle \propto \alpha/p^4$, which could give rise to potential IR divergences when performing explicit loop calculations. Nevertheless, it is possible to introduce an IR regularizing mass term for the Stueckelberg field compatible with the local BRST invariance. Indeed, from the transformation of the Stueckelberg field given in eq. \eqref{eqsxi}, it can be checked that
\begin{equation}
s_{loc}\left(\frac{\xi^a\xi^a}{2}\right)=-\xi^a c^a \,.    \label{prost}
\end{equation}
This interesting property of the Stueckelberg field can be derived from eq.\eqref{mbrst}, {\it  i.e.}
\begin{equation}
s_{loc} (e^{ig\xi}) = - ig c e^{ig\xi} \;.  \label{ig1}
\end{equation}
Expanding the exponential in Taylor series, one gets
\begin{equation}
s_{loc} \left( 1 + ig \xi - \frac{g^2}{2} \xi \xi - i \frac{g^3}{3!} \xi \xi \xi + \cdot  \cdot  \right)  = -igc \left(   1 + ig \xi - \frac{g^2}{2} \xi \xi - i \frac{g^3}{3!} \xi \xi \xi + \cdot  \cdot \right)  \;. \label{ig2}
\end{equation}
Multiplying both sides of eq.\eqref{ig2} by $\xi$, yields
\begin{equation}
\xi \; s_{loc} \left( 1 + ig \xi - \frac{g^2}{2} \xi \xi - i \frac{g^3}{3!} \xi \xi \xi + \cdot  \cdot  \right)  = -ig \xi \;c \left(   1 + ig \xi - \frac{g^2}{2} \xi \xi - i \frac{g^3}{3!} \xi \xi \xi + \cdot  \cdot \right)  \;. \label{ig3}
\end{equation}
Equating order by order in $g$ the expression \eqref{ig3} immediately provides eq.\eqref{prost} at leading order.

\noindent Due to equation \eqref{prost}, we can introduce the following BRST-exact term
\begin{equation}
 S_{IRR}=\int d^4x \frac{1}{2} s_{loc}\left(\rho\xi^a\xi^a\right) =\int d^4x \left( \frac{1}{2}M^4\xi^a\xi^a+\rho\xi^ac^a  \right) \,,
\label{SIRR}
\end{equation}
where $(\rho, M)$ are constant parameters transforming as
 \begin{equation}
 s_{loc}\rho=M^{4}\,,\qquad s_{loc}M^{4}=0\,.
 \end{equation}
The parameter $\rho$ has ghost number $-1$, while $M$ has ghost number zero. As it is apparent from equation \eqref{SIRR}, the term $S_{IRR}$ provides an IR regularization for the Stueckelberg field in a BRST-invariant way, yielding in fact a propagator for the Stueckelberg field which behaves now like $\langle \xi \xi \rangle\propto \frac{\alpha}{p^4+\alpha M^4}$.  At the very end of the computation of the correlation functions the parameters $(M,\rho)$ will be set to zero.

Before ending this section, it might be worth to spend a few words on an important aspect displayed by the model in the Landau gauge, {\it i.e.} $\partial_\mu A_\mu=0$, which corresponds to $\alpha=0$. As one checks out from Appendix \ref{AppProps}, all propagators of all fields are IR safe in the Landau gauge. In other words, when $\alpha=0$, the introduction of the regularizing infrared mass $M$ is redundant. In particular, from Appendix \ref{AppProps}, one sees that
$ \langle \xi \xi \rangle_{Landau} = \langle A_\mu \xi \rangle_{Landau} = \langle A_\mu \tau \rangle_{Landau} =0 $, while all other two-points correlation functions are IR safe. This is an important property of the Landau gauge, which expresses in terms of Feynman rules the decoupling of the Stueckelberg field, reflecting the expected fact that, when $\partial_\mu A_\mu=0$, the higher order terms of the infinite series \eqref{hhh3g}  become harmless, due to the presence of the divergence $\partial_\mu A_\mu$. From this useful feature of the Landau gauge one infers that the existence of the limit $M^2 \rightarrow 0$ is apparent  for correlation functions $\langle O(x) O(y) \rangle$ of BRST invariant composite operators $O(x)$. In fact, as we shall show in details in the next section, the existence of an exact BRST symmetry will enable us to prove that $\langle O(x) O(y) \rangle$ turns out to be independent from the gauge parameter $\alpha$ to all orders. As a consequence, the correlation function $\langle O(x) O(y) \rangle$ can be directly evaluated in the Landau gauge, $\alpha=0$, which is IR safe, due to the aforementioned decoupling properties of the Stueckelberg field $\xi$.

% % % % % % % % % % % % % % % % % % % % % % % % % % % % % % % % %

\subsection{$i$-particles}

In order to elucidate the nature of the fundamental excitations of the theory, we look at the quadratic part of the BRST singlet sector of the action \eqref{gzlocalhorizon3}, which, bearing in mind that $F^2$ is gauge invariant, i.e.
\begin{equation}
F_{\mu\nu}^a(A)F_{\mu\nu}^a(A)=F_{\mu\nu}^a(A^h)F_{\mu\nu}^a(A^h)\,,
\end{equation}
is given by
\begin{eqnarray}
 S_{quad}^{singlet} =\int\,d^4x \left(\frac{1}{2}(A^h)^a_\mu (-\partial^2) (A^h)^a_\mu -\bar\lambda^{ac}_\mu  (-\partial^2) \lambda^{ac}_\mu + \bar\zeta^{ac}_\mu (-\partial^2)\zeta^{ac}_\mu
+ g\gamma^2\,f^{abc}(A^h)^a_\mu(\lambda_\mu^{bc} + \bar\lambda^{bc}_\mu)\right),\nonumber\\
\label{gzlocalhorizon-quadphys}
\end{eqnarray}
where we have already taken into account the fact that the multiplier $\tau^a$ enforces the transversality of  $(A^h)^a_\mu$.
Defining
\begin{eqnarray}
\lambda^{ab}_{\mu} = \frac{1}{\sqrt{2}}\left( V^{ab}_{\mu} - iU_\mu^{ab} \right) \,,
 \label{VU}
\end{eqnarray}
we  have
\begin{eqnarray}
S_{quad}^{singlet}&=&\int\,d^4x \left(\frac{1}{2}(A^h)^a_\mu (-\partial^2) (A^h)^a_\mu - \frac{1}{2} V^{ac}_\mu (-\partial^2) V^{ac}_\mu\right.\nonumber\\ &&\left.-  \frac{1}{2} U^{ac}_\mu (-\partial^2) U^{ac}_\mu +\bar\zeta^{ac}_\mu(-\partial^2)\zeta^{ac}_\mu
+ g\gamma^2\sqrt{2}\,f^{abc}(A^h)^a_\mu V_\mu^{bc} \right)\,.
\label{gzlocalhorizon-quadphys2}
\end{eqnarray}
 Expression \eqref{gzlocalhorizon-quadphys2} can be diagonalized in the following way: for  SU($N$), we decompose the field $V^{ab}$ as
\begin{eqnarray}
V^{ab}_{\mu} =\frac{1}{\sqrt{N}} f^{abc} V_\mu^{c} + S_\mu^{ab} \,,
 \label{V2}
\end{eqnarray}
where
\begin{eqnarray}
V^{a}_{\mu} \equiv \frac{1}{\sqrt{N}} f^{abc} V_\mu^{bc} \,,
 \label{V}
\end{eqnarray}
and
\begin{eqnarray}
 f^{abc} S_\mu^{ab} =0 \,.
 \label{S}
\end{eqnarray}
The validity of equation \eqref{V2} can be  easily checked with the help of
\begin{eqnarray}
 f^{abc} f^{abd} =N \delta^{cd} \,.
 \label{f}
\end{eqnarray}
The action becomes now
\begin{eqnarray}
 S_{quad}^{singlet} &=&\int\,d^4x \left(\frac{1}{2}(A^h)^a_\mu (-\partial^2) (A^h)^a_\mu - \frac{1}{2} V^{a}_\mu (-\partial^2) V^{a}_\mu
+ g\sqrt{2N}\gamma^2\,(A^h)^a_\mu V_\mu^{a} \right)\nonumber\\
&& + \int\,d^4x \left(- \frac{1}{2} S^{ac}_\mu (-\partial^2) S^{ac}_\mu -  \frac{1}{2} U^{ac}_\mu (-\partial^2) U^{ac}_\mu +\bar\zeta^{ac}_\mu(-\partial^2)\zeta^{ac}_\mu \right)\,.
\label{gzlocalhorizon-quadphys3}
\end{eqnarray}
Therefore, introducing the complex fields
\begin{eqnarray}
\eta^{a}_{\mu} = (A^h)^a_\mu + i V_\mu^{a} \,,  \qquad {\bar \eta}^a_\mu = (A^h)^a_\mu - i V_\mu^{a}
 \label{ipart}
\end{eqnarray}
we obtain
\begin{eqnarray}
S_{quad}^{singlet} &=& \int\,d^4x \left(\frac{1}{4}\eta^a_\mu \left(-\partial^2 -ig\sqrt{2N}\gamma^2\right) \eta^a_\mu + \frac{1}{4}\bar\eta^a_\mu \left(-\partial^2 +ig\sqrt{2N}\gamma^2\right) \bar\eta^a_\mu \right)\nonumber\\
&& + \int\,d^4x \left(- \frac{1}{2} S^{ac}_\mu (-\partial^2) S^{ac}_\mu -  \frac{1}{2} U^{ac}_\mu (-\partial^2) U^{ac}_\mu +\bar\zeta^{ac}_\mu(-\partial^2)\zeta^{ac}_\mu \right).
\label{gzlocalhorizon-ipart}
\end{eqnarray}
As it is apparent from expression \eqref{gzlocalhorizon-ipart}, the fields $(\eta, \bar \eta)$ correspond to a pair of unphysical excitations, called $i$-particles \cite{Baulieu:2009ha}, with  complex conjugate squared masses $\pm ig\sqrt{N}\gamma^2$.  As discussed in details in \cite{Baulieu:2009ha}, this kind of excitation is suitable to describe confined degrees of freedom. Moreover, it turns out that physical bound states operators displaying a K\"all\'en-Lehmann spectral representation can be constructed by combining  pairs of $i$-particles, see  \cite{Baulieu:2009ha}. Examples of such composite operators are
\begin{eqnarray}
{\cal O}^{(1)} &=& \bar{\eta}_{\mu\nu}\eta_{\mu\nu}\nonumber\\
{\cal O}^{(2)} &=& \varepsilon^{\mu\nu\rho\sigma}\bar{\eta}_{\mu\nu}\eta_{\rho\sigma}  \;,
\label{composite-ipart}
\end{eqnarray}
where
\begin{eqnarray}
\eta_{\mu\nu} = \partial_{\mu}\eta_{\nu} - \partial_{\nu}\eta_{\mu}\,.
\label{fieldstrenght-ipart}
\end{eqnarray}
It is remarkable that the fields $(\eta, \bar \eta)$ corresponding to the $i$-particles are BRST singlets. As a consequence, the composite operators \eqref{composite-ipart} are BRST singlets too.

%%%%%%%%%%%%%%%%%%
\section{Ward identities and their consequences}\label{section:Ward}

Having achieved a local formulation of the BRST transformations, we can proceed with the derivation of the Ward identities and with the analysis of their consequences. To that purpose we employ the powerful trick of extending the BRST transformations on the gauge parameter $\alpha$, see  \cite{Piguet:1995er,Piguet:1984js}, {\it i.e.}
\begin{equation}
s_{loc} \alpha = \chi \;, \qquad s_{loc} \chi = 0 \;, \label{extbrst}
\end{equation}
where $\chi$ is a parameter with ghost number 1, which will be set to zero at the end. As explained in  \cite{Piguet:1995er,Piguet:1984js}, the extended BRST transformations, eqs.\eqref{brstgammalocal}, \eqref{extbrst}, will enable us to prove in a purely algebraic way that the correlation functions of local operators belonging to the cohomology of the BRST operator $s_{loc}$ are independent from the gauge parameter $\alpha$.

Taking into account the extended BRST transformation \eqref{extbrst}, the gauge fixing term becomes now
\begin{equation}
s_{loc} \int d^4x \left( -i \frac{\alpha}{2} {\bar c}^a b^a + {\bar c}^a \partial_\mu A^a_\mu \right) = \int d^4x \left( \frac{\alpha}{2} b^a b^a + i b^a  \partial_\mu A^a_\mu - i \frac{\chi}{2} {\bar c}^a b^a + {\bar c}^a \partial_\mu D_\mu^{ab}(A) c^b    \right)  \;, \label{ngf}
\end{equation}
so that the action \eqref{gzlocalhorizon2} reads
\begin{eqnarray}
S_{GZ}^{LCG} &=& S_{YM} + \int d^4x \left( \alpha\frac{b^ab^a}{2} + ib^a \partial_\mu A_\mu^a - i \frac{\chi}{2} {\bar c}^a b^a + {\bar c}^a \partial_\mu D_\mu^{ab}(A) c^b \right) +\int d^4x \;\tau^{a}\,\partial_{\mu}(A^h)^{a}_{\mu}
    \;\nonumber\\
&& \hspace{1cm} + \int\,d^4x \left(-\bar\varphi^{ac}_\mu{\cal M}(A^h)^{ab}\varphi^{bc}_\mu + \bar\omega^{ac}_\mu{\cal M}(A^h)^{ab}\omega^{bc}_\mu
+ g\frac{\gamma^2}{\sqrt{2}}\,f^{abc}(A^h)^a_\mu(\varphi_\mu^{bc} + \bar\varphi^{bc}_\mu)\right)
\nonumber\\
&& \hspace{1cm} +\int\,d^4x \left(-\bar\beta^{ac}_\mu{\cal M}(A^h)^{ab}\beta^{bc}_\mu + \bar\zeta^{ac}_\mu{\cal M}(A^h)^{ab}\zeta^{bc}_\mu
- g\frac{\gamma^2}{\sqrt{2}}\,f^{abc}(A^h)^a_\mu(\beta_\mu^{bc} + \bar\beta^{bc}_\mu)\right) \;.
\label{gzlocalhorizon2b}
\end{eqnarray}
We are now ready to establish the Ward identities of the theory. Following the general procedure of the algebraic renormalization \cite{Piguet:1995er}, we introduce a set of BRST-invariant external sources $(\Omega^a_\mu, L^a, K^a)$ coupled to the non-linear BRST variations of the elementary fields, namely we start with the complete classical action
\begin{eqnarray}
\Sigma&=&S_{GZ}^{LCG}+S_{IRR} + \int d^{4}x\left(\Omega^{a}_{\mu}\,s_{loc}A^{a}_{\mu}+L^{a}\,s_{loc}c^{a}+K^{a}\,s_{loc}\xi^{a}\right) \nonumber\\
&=&S_{GZ}^{LCG}+\int d^{4}x\left(\Omega^{a}_{\mu}\,s_{loc}A^{a}_{\mu}+L^{a}\,s_{loc}c^{a}+K^{a}\,s_{loc}\xi^{a}\right)+\int d^{4}x\left(\frac{M^{4}}{2}\,\xi^{a}\xi^{a}+\rho\,\xi^{a}c^{a}\right)\,,
\end{eqnarray}
where
\begin{equation}
s_{loc}\Sigma=0\,,\qquad s_{loc}s_{loc}=0\,,
\end{equation}
with $s_{loc}$ being the local version of the nonperturbative BRST,  eqs.\eqref{brstgammalocal}, \eqref{extbrst}. The complete action $\Sigma$ turns out to obey the following Ward identities \cite{Piguet:1995er}:
\begin{itemize}
\item{The Slavnov-Taylor identity:
\begin{eqnarray}
\mathcal{S}(\Sigma)&=&\int d^{4}x\bigg[
\frac{\delta\Sigma}{\delta\Omega^{a}_{\mu}}\frac{\delta\Sigma}{\delta A^{a}_{\mu}}
+\frac{\delta\Sigma}{\delta L^{a}}\frac{\delta\Sigma}{\delta c^{a}}
+\frac{\delta\Sigma}{\delta K^{a}}\frac{\delta\Sigma}{\delta \xi^{a}}
+ib^{a}\,\frac{\delta\Sigma}{\delta\bar{c}^{a}}
+\omega^{ab}_{\mu}\,\frac{\delta\Sigma}{\delta \varphi^{ab}_{\mu}}
+ \omega^{ab}_{\mu}\,\frac{\delta\Sigma}{\delta {\beta}^{ab}_{\mu}}
\nonumber\\
&&
+ (\bar\varphi^{ab}_{\mu}+{\bar \beta}^{ab}_{\mu})\,\frac{\delta\Sigma}{\delta\bar\omega^{ab}_{\mu}}\bigg]
+M^{4}\,\frac{\partial\Sigma}{\partial \rho}
+\chi\,\frac{\partial\Sigma}{\partial\alpha}=0\,.    \label{stc}
\end{eqnarray}}
\item{The gauge-fixing equation:
\begin{equation}
\frac{\delta\Sigma}{\delta b^{a}}=i\,\partial_{\mu}A^{a}_{\mu}+\alpha\, b^{a}-\frac{i}{2}\,\chi\,\bar{c}^{a}\,.    \label{gfc}
\end{equation}}
\item{The anti-ghost equation:
\begin{equation}
\frac{\delta\Sigma}{\delta\bar{c}^{a}}+\partial_{\mu}\frac{\delta\Sigma}{\delta\Omega^{a}_{\mu}}=\frac{i}{2}\,\chi\,b^{a}\,.   \label{gec}
\end{equation}}
\end{itemize}
 To exploit the content of these identities at the quantum level, we introduce the $1PI$ generating functional $\Gamma$
\begin{equation}
\Gamma=\Sigma +\sum_{n=1}^{\infty}\hbar^{n}\Gamma^{(n)}\,,
\end{equation}
and write down the quantum version of the identities \eqref{stc}, \eqref{gfc}, \eqref{gec}, {\it i.e.}
\begin{eqnarray}
\mathcal{S}(\Gamma)&=&\int d^{4}x\bigg[
\frac{\delta\Gamma}{\delta\Omega^{a}_{\mu}}\frac{\delta\Gamma}{\delta A^{a}_{\mu}}
+\frac{\delta\Gamma}{\delta L^{a}}\frac{\delta\Gamma}{\delta c^{a}}
+\frac{\delta\Gamma}{\delta K^{a}}\frac{\delta\Gamma}{\delta \xi^{a}}
+ib^{a}\,\frac{\delta\Gamma}{\delta\bar{c}^{a}}
+\omega^{ab}_{\mu}\,\frac{\delta\Gamma}{\delta \varphi^{ab}_{\mu}}
+ \omega^{ab}_{\mu}\,\frac{\delta\Gamma}{\delta {\beta}^{ab}_{\mu}}
\nonumber\\
&&
+ (\bar\varphi^{ab}_{\mu}+{\bar \beta}^{ab}_{\mu})\,\frac{\delta\Gamma}{\delta\bar\omega^{ab}_{\mu}}\bigg]
+M^{4}\,\frac{\partial\Gamma}{\partial \rho}
+\chi\,\frac{\partial\Gamma}{\partial\alpha}=0\,.    \label{stq}
\end{eqnarray}
\begin{equation}
\frac{\delta\Gamma}{\delta b^{a}}=i\,\partial_{\mu}A^{a}_{\mu}+\alpha\, b^{a}-\frac{i}{2}\,\chi\,\bar{c}^{a}\,.   \label{gfq}
\end{equation}
\begin{equation}
\frac{\delta\Gamma}{\delta\bar{c}^{a}}+\partial_{\mu}\frac{\delta\Gamma}{\delta\Omega^{a}_{\mu}}=\frac{i}{2}\,\chi\,b^{a}\,.    \label{geq}
\end{equation}

We move then from the $1PI$ correlation functions to the connected ones by means of the Legendre transformation
\begin{equation}
Z[J,\mathcal{J},\mu]=\Gamma[\Phi,\mathcal{J},\mu]+\sum_{i}\int d^{4}x J^{(\Phi)}_{i}\Phi_{i}\,,    \label{zj}
\end{equation}
 whereby $J$ stands for the standard sources coupled to the fields of the theory and
\begin{eqnarray}
\Phi_i&\equiv&\{A,b,c,\bar{c},\xi, \tau,\varphi,\bar\varphi,\omega,\bar\omega,\beta,\bar\beta,\zeta,\bar{\zeta} \}\,,\nonumber\\
\mathcal{J}&\equiv&\{\Omega,L,K\}\,,\nonumber\\
\mu&\equiv&\{\rho,M^{4},\alpha,\chi\}\,.
\end{eqnarray}
From expression \eqref{zj} we have
\begin{equation}
\begin{tabular}{rclrcl}
$\displaystyle \frac{\delta\Gamma}{\delta\Phi^{bos}_{i}}$&$=$&$-J^{(\Phi^{bos})}_{i}\,,\qquad$&$\displaystyle\frac{\delta Z}{\delta J^{(\Phi^{bos})}_{i}}$&$=$&$\Phi^{bos}_{i}\,,$
\end{tabular}
\end{equation}
for bosonic fields and
\begin{equation}
\begin{tabular}{rclrcl}
$\displaystyle \frac{\delta\Gamma}{\delta\Phi^{fer}_{i}}$&$=$&$J^{(\Phi^{fer})}_{i}\,,\qquad$&$\displaystyle\frac{\delta Z}{\delta J^{(\Phi^{fer})}_{i}}$&$=$&$\Phi^{fer}_{i}\,,$
\end{tabular}
\end{equation}
for fermionic fields, so that
\begin{equation}
\begin{tabular}{rclrcl}
$\displaystyle \frac{\delta\Gamma}{\delta\mathcal{J}}$&$=$&$\displaystyle \frac{\delta Z}{\delta\mathcal{J}}\,,\qquad$
&$\displaystyle \frac{\partial\Gamma}{\partial\mu}$&$=$&$\displaystyle \frac{\partial Z}{\partial\mu}\,.$
\end{tabular}
\end{equation}
When written in terms of the generating functional $Z[J,\mathcal{J},\mu]$, the previous Ward identities take the form:
\begin{itemize}
\item{The Slavnov-Taylor identity:
\begin{eqnarray}
&&\int d^{4}z\Bigg[
-J^{(A)a}_{\mu}(z)\frac{\delta Z}{\delta\Omega^{a}_{\mu}(z)}
+J^{(c)a}(z)\frac{\delta Z}{\delta L^{a}(z)}
-J^{(\xi)a}(z)\frac{\delta Z}{\delta K^{a}(z)}
+iJ^{(\bar{c})a}(z)\frac{\delta Z}{\delta J^{(b)a}(z)}
\nonumber\\
&&
- (J^{(\varphi)ab}_{\mu}(z)+J^{(\beta)ab}_{\mu}(z))\frac{\delta Z}{\delta J^{(\omega)ab}_{\mu}(z)}
+  J^{(\bar\omega)ab}_{\mu}(z)\left(\frac{\delta Z}{\delta J^{(\bar\varphi)ab}_{\mu}(z)}+\frac{\delta Z}{\delta J^{({\bar \beta})ab}_{\mu}(z)}\right)\Bigg]
\nonumber\\
&&
+M^{4}\,\frac{\partial Z}{\partial \rho}
+\chi\,\frac{\partial Z}{\partial \alpha}=0\,.
\label{ST_Z}
\end{eqnarray}}
\item{The gauge-fixing equation:
\begin{equation}
-J^{(b)a}(x)=i\,\partial^{x}_{\mu}\frac{\delta Z}{\delta J^{(A)a}_{\mu}(x)}+\alpha\frac{\delta Z}{\delta J^{(b)a}(x)}
-\frac{i}{2}\,\chi\,\frac{\delta Z}{\delta J^{(\bar{c})a}(x)}.
\label{gauge_fixing_Z}
\end{equation}}
\item{The anti-ghost equation:
\begin{equation}
J^{(\bar{c})a}(x)+\partial^{x}_{\mu}\frac{\delta Z}{\delta\Omega^{a}_{\mu}(x)}=\frac{i}{2}\,\chi\,\frac{\delta Z}{\delta J^{(b)a}(x)}\,.
\label{agheq}
\end{equation}}
\end{itemize}

In the following, we shall explore some implications of these identities for the two-point gluon correlation
function and to the correlation functions of BRST-invariant operators.

%%%%%%%%%%%%%%%
\subsection{The longitudinal part of the gluon propagator}

The first important consequence stemming from the identities \eqref{ST_Z}, \eqref{gauge_fixing_Z} and \eqref{agheq} is that they completely determine the longitudinal part of the propagator of the gauge field $A^a_\mu$.

Acting with the test operator
\begin{equation}
\frac{\delta}{\delta J^{(A)a}_{\mu}(x)}\frac{\delta}{\delta J^{(\bar{c})b}(y)}
\end{equation}
on the Slavnov-Taylor identity, eq.\eqref{ST_Z}, and setting all sources and parameters $M$ and $\chi$ equal to zero, one gets
\begin{equation}
\frac{\delta}{\delta J^{(\bar{c})b}(y)}\frac{\delta Z}{\delta \Omega^{a}_{\mu}(x)}-i\frac{\delta^{2} Z}{\delta J^{(b)b}(y)\delta J^{(A)a}_{\mu}(x)}=0\,.
\label{ST_after_test}
\end{equation}
Let us now act on eq.\eqref{ST_after_test} with the partial derivative $\partial^{x}_{\mu}$, obtaining
\begin{equation}
\frac{\delta}{\delta J^{(\bar{c})b}(y)}\partial^{x}_{\mu}\frac{\delta Z}{\delta \Omega^{a}_{\mu}(x)}-i\partial^{x}_{\mu}\frac{\delta^{2} Z}{\delta J^{(b)b}(y)\delta J^{(A)a}_{\mu}(x)}=0\,.   \label{jja}
\end{equation}
Making use of the anti-ghost equation \eqref{agheq} with $\chi$ equal to zero, it follows that eq.\eqref{jja} yields
\begin{equation}
\frac{\delta}{\delta J^{(\bar{c})b}(y)}J^{(\bar{c})a}(x)
+i\partial^{x}_{\mu}\frac{\delta^{2}Z}{\delta J^{(b)b}(y)\delta J^{(A)a}_{\mu}(x)}=0
\end{equation}
or
\begin{equation}
\delta^{ab}\delta(x-y)
+i\partial^{x}_{\mu}\frac{\delta^{2}Z}{\delta J^{(A)a}_{\mu}(x)\delta J^{(b)b}(y)}=0\,,
\end{equation}
which, in momentum space, becomes
\begin{equation}
p_{\mu}\langle A_\mu^a \;b^b \rangle_{\mathrm{conn}}(p)=-\delta^{ab}\,,
\end{equation}
where we have defined $\langle A_\mu^a \;b^b \rangle_{\mathrm{conn}}(p)$ in such a way that
\begin{equation}
\frac{\delta^{2}Z}{\delta J^{(A)a}_{\mu}(x)\delta J^{(b)b}(y)}=-\langle A^{a}_{\mu}(x)b^{b}(y)\rangle_{\mathrm{conn}}=-\int \frac{d^{4}p}{(2\pi)^{4}}\,\langle A_\mu^a \;b^b\rangle_{\mathrm{conn}}(p)\,e^{ip(x-y)}\,.
\end{equation}
Therefore, from Lorentz\footnote{ Or more precisely, Euclidean $4D$ rotational invariance in our case.} invariance, it follows that
\begin{equation}
\langle A_\mu^a \;b^b \rangle_{\mathrm{conn}}(p)=-\frac{p_{\mu}}{p^{2}}\,\delta^{ab}\,.
\end{equation}
It remains now to apply the test operator $\frac{\delta}{\delta J^{(A)b}_{\nu}(y)}$ on eq.\eqref{gauge_fixing_Z} and set all sources and $\chi$ equal to zero, obtaining
\begin{equation}
i\partial^{x}_{\mu}\frac{\delta^{2}Z}{\delta J^{(A)a}_{\mu}(x)\delta J^{(A)b}_{\nu}(y)}+\alpha\,\frac{\delta^{2}Z}{\delta J^{(A)b}_{\nu}(y)\delta J^{(b)a}(x)}=0\,,
\end{equation}
which gives, in momentum space,
\begin{equation}
-p_{\mu}\langle A_\mu^a \; A_\nu^b  \rangle_{\mathrm{conn}}(p)+\alpha\,\langle A_\nu^b \; b^a\rangle_{\mathrm{conn}}(-p)=0\,,
\end{equation}
or
\begin{equation}
p_{\mu}\langle A_\mu^a \;A_\nu^b \rangle_{\mathrm{conn}}(p)=\alpha\,\frac{p_{\nu}}{p^{2}}\,\delta^{ab}\,.
\end{equation}
Therefore, decomposing the gauge propagator $\langle A^a_\mu\; A^b_\nu \rangle_{\mathrm{conn}}$ into transverse and longitudinal components
\begin{equation}
\langle A_\mu^a \; A_\nu^b \rangle_{\mathrm{conn}}(p)=\left[\left(\delta_{\mu\nu}-\frac{p_{\mu}p_{\nu}}{p^{2}}\right)\,D(p^{2})+\frac{p_{\mu}p_{\nu}}{p^{2}}\,G(p^{2})\right]\,\delta^{ab} \label{decgp}
\end{equation}
it follows that the longitudinal component $G(p^{2})$ is completely determined to all orders to be
\begin{equation}
G(p^2)=\frac{\alpha}{p^{2}}  \;,    \label{G}
\end{equation}
a result which is in full agreement with the recent lattice data \cite{Cucchieri:2009kk,Bicudo:2015rma}  as well as with the results following from the analysis of the Dyson-Schwinger  equations \cite{Aguilar:2015nqa,Huber:2015ria,Aguilar:2016vin}.  The linear covariant gauge was also analyzed in \cite{Siringo:2015gia,Siringo:2015wtx}.  In \cite{Capri:2015nzw}, an alternative proof of this fact was presented, directly  at the level of the path integral. It is worth mentioning that the result \eqref{G} is hard to show if one would not have an exact BRST symmetry for the Gribov-Zwanziger action at one's disposal.

%%%%
\subsection{Gauge-parameter independence of correlation functions of BRST-invariant composite operators}

A second relevant consequence which follows directly from the Slavnov-Taylor identity, eq.\eqref{ST_Z}, is the independence from the gauge parameter $\alpha$ of the correlation functions of composite operators which belong to the cohomology of the BRST operator $s_{loc}$. To establish this property we follow the procedure outlined in \cite{Piguet:1995er}.

Let $O(x)$ denote a local composite operator with vanishing ghost number and belonging to the cohomolgy of $s_{loc}$, namely
\begin{equation}
s_{loc} O = 0 \;, \qquad O \neq  s_{loc} {\tilde O} \;, \label{oop}
\end{equation}
for any local operator ${\tilde O}$. The correlation functions $\left\langle {\cal O}(x_{1}) \cdots {\cal O}(x_{n}) \right\rangle_{\mathrm{conn}}$ are obtained by adding to the starting action the BRST-invariant term $\int d^4x \left( J^{(\cal O)}(x)\; O(x) \right)$, with $J^{(\cal O)}$ being an external invariant  source. Due to the BRST invariance of $\int d^4x \left( J^{(\cal O)}(x) \; O(x) \right) $, the Slavnov-Taylor identity,  eq.\eqref{ST_Z}, remains unmodified.

The correlators  $\left\langle {\cal O}(x_{1}) \cdots {\cal O}(x_{n}) \right\rangle_{\mathrm{conn}}$  are derived  by differentiating the generating function $Z$ with respect to $J^{(\cal O)}$ and by setting all sources and parameters $(M^4, \rho, \chi)$ equal to zero, {\it i.e.}
\begin{eqnarray}
\left\langle {\cal O}(x_{1}) \cdots {\cal O}(x_{n}) \right\rangle_{\mathrm{conn}} = \frac{\delta }{\delta J^{(\cal O)}(x_{n})} \cdots \frac{\delta}{\delta J^{(\cal O)}(x_{1})} Z {\Big|}_{J=\mathcal{J}=0, M=\rho=\chi=0}\;.   \label{defg}
\end{eqnarray}
To prove the gauge-parameter independence of the correlation functions $\left\langle {\cal O}(x_{1}) \cdots {\cal O}(x_{n}) \right\rangle_{\mathrm{conn}}$, we act with the test operator
\begin{equation}
\frac{\delta }{\delta J^{(\cal O)}(x_{n})} \cdots \frac{\delta}{\delta J^{(\cal O)}(x_{1})}
\end{equation}
on the Slavnov-Taylor identity, eq.\eqref{ST_Z}, and afterwards we derive with respect to the parameter $\chi$, yielding the equation
\begin{eqnarray}
\label{gaugeindp1}
 \frac{\partial}{\partial\alpha}\left[\frac{\delta^{n} Z}{\delta J^{(\cal O)}(x_{n}) \cdots \delta J^{(\cal O)}(x_{1})}{\Big|}_{J=\mathcal{J}=0, M=\rho=0}\right]  - \chi \frac{\partial^2}{\partial\chi \partial\alpha }\left[\frac{\delta^{n} Z}{\delta J^{(\cal O)}(x_{n}) \cdots \delta J^{(\cal O)}(x_{1})}{\Big|}_{J=\mathcal{J}=0, M=\rho=0 }\right]= 0\;.
\end{eqnarray}
Setting now the parameter $\chi$ to zero, we finally get
\begin{eqnarray}
\label{gaugeindp}
\frac{\partial }{\partial \alpha} \left\langle {\cal O}(x_{1}) \cdots {\cal O}(x_{n}) \right\rangle_{\mathrm{conn}}  = 0\;,
\end{eqnarray}
which establishes the all-order independence of the correlation functions $\left\langle {\cal O}(x_{1}) \cdots {\cal O}(x_{n}) \right\rangle_{\mathrm{conn}}$ from the gauge parameter $\alpha$.  In particular, the vacuum energy (zero point function) is $\alpha$-independent, and as a consequence also the Gribov parameter $\gamma$ which follows from extremizing the vacuum energy.

%%%%%%%%%%%%%%%%%%
\section{Dimension-two condensates and construction of the Refined GZ action}\label{section:RGZ}

It is known that the GZ  vacuum  exhibits the formation of dimension-two condensates which are energetically favoured, giving rise to a refined action called the RGZ action, see \cite{Dudal:2007cw,Dudal:2008sp,Dudal:2011gd}. In the present BRST-invariant formulation of the linear covariant gauges, these dimension-two condensates are easily identified, the corresponding dimension-two operators being given by
\begin{equation}
(A^h)_{\mu}^a(A^h)_{\mu}^a   \;, \qquad    \left(\bar\omega^{ab}_{\mu}\omega^{ab}_{\mu} - \bar\varphi^{ab}_{\mu}\varphi^{ab}_{\mu} - \bar\beta^{ab}_{\mu}\beta^{ab}_{\mu}
+ \bar\zeta^{ab}_{\mu}\zeta^{ab}_{\mu}  \right)  \;.    \label{cond}
\end{equation}
We underline that both operators appearing in eq. \eqref{cond} belong to the cohomology of the BRST operator $s_{loc}$, as it is easily  checked. Taking into account these dimension two operators,  for the BRST-invariant RGZ action we get
%\eqref{gzh2}
\begin{eqnarray}
 S_{RGZ}^{LCG} &=& S_{YM} + S_{cond} + S_{IRR} + \int d^4x \left( \alpha\frac{b^ab^a}{2} + ib^a \partial_\mu A_\mu^a + {\bar c}^a \partial_\mu D_\mu^{ab}(A) c^b \right) +\int d^4x \;\tau^{a}\,\partial_{\mu}(A^h)^{a}_{\mu}.
    \;\nonumber\\
&& \hspace{1cm} + \int\,d^4x \left(-\bar\varphi^{ac}_\mu{\cal M}(A^h)^{ab}\varphi^{bc}_\mu + \bar\omega^{ac}_\mu{\cal M}(A^h)^{ab}\omega^{bc}_\mu
+ g\frac{\gamma^2}{\sqrt{2}}\,f^{abc}(A^h)^a_\mu(\varphi_\mu^{bc} + \bar\varphi^{bc}_\mu)\right),
\nonumber\\
&& \hspace{1cm} +\int\,d^4x \left(-\bar\beta^{ac}_\mu{\cal M}(A^h)^{ab}\beta^{bc}_\mu + \bar\zeta^{ac}_\mu{\cal M}(A^h)^{ab}\zeta^{bc}_\mu
- g\frac{\gamma^2}{\sqrt{2}}\,f^{abc}(A^h)^a_\mu(\beta_\mu^{bc} + \bar\beta^{bc}_\mu)\right)  \;,
\label{gzlocalhorizon22}
\end{eqnarray}
with $S_{IRR}$ given by expression \eqref{SIRR} and  \footnote{As pointed out in \cite{Dudal:2008sp} before, $\mu^2$ should be positive to avoid tachyonic modes in the $\bar{\omega}\omega$ sector.}
\begin{equation}
S_{cond}=\int d^4x\left\{\frac{m^2}{2}(A^h)_{\mu}^a(A^h)_{\mu}^a
+\mu^2\left(\bar\omega^{ab}_{\mu}\omega^{ab}_{\mu} - \bar\varphi^{ab}_{\mu}\varphi^{ab}_{\mu} - \bar\beta^{ab}_{\mu}\beta^{ab}_{\mu}
+  \bar\zeta^{ab}_{\mu}\zeta^{ab}_{\mu}  \right)\right\} \;.    \label{condact}
\end{equation}
 We note that in the auxiliary fields sector, a single mass parameter is used. If different parameters were used, we could simply fix then to be equal (up to a sign) by demanding that $S_{cond}$ is invariant under the non-perturbative BRST transformations and the discrete transformations \eqref{discretesymm}. The parameters $(m^2, \mu^2)$ in expression \eqref{condact} are dynamical parameters which, as much as the Gribov parameter $\gamma^2$,  can be determined order by order through the evaluation of the effective potential for the dimension two operators of eq.\eqref{cond}, see \cite{Dudal:2011gd}.  Notice also that, as in the case of $\gamma^2$, the parameters $(m^2, \mu^2)$ are coupled to the operators \eqref{cond}, which belong to the cohomology of the BRST operator. As such, $(m^2, \mu^2)$ are physical parameters of the theory which will enter the physical correlation functions of BRST invariant operators.

%%%%%%%%%%%%%%%%%%%%%%%%%%%%%%

\section{Inclusion of matter fields}\label{section:matter}

It is interesting to point out that this fully local BRST-invariant description is not restricted to the case of pure-gauge theories.
Gauge-invariant matter fields,   analogous to $A^h$,  may be constructed for various fields belonging to different representations of
the SU($N$) color group, allowing for the modeling \footnote{ It would remain, in the future, to work out a self-consistent dynamical realization of such model, starting from the standard Yang-Mills matter actions.} of nonperturbative gauge-interacting matter. In what follows, we discuss two particular cases of
phenomenological interest: adjoint scalar fields and Dirac fermions in the fundamental representation.

\subsection{Scalar fields in the adjoint representation}

Consider adjoint scalar fields
\begin{equation}
 \phi = \phi^a T^a  \;,
\end{equation}
whose BRST transformation is
\begin{equation}
s \phi = ig [\phi, c]  \;.
\end{equation}
A gauge invariant scalar field is obtained by making use of the Stueckelberg field $\xi$ (defined in eq. \eqref{hxi}), as
\begin{equation}
 \phi^h = h^{\dagger} \phi h  %\;, \qquad h=e^{ig\,\xi^{a}T^{a}}
 \;.
\end{equation}
It is easy to verify that $\phi^h$ is left invariant by the BRST transformations, {\it i.e.}
\begin{equation}
 s \phi^h = 0\,.
\end{equation}
Generalizing the construction outlined in \cite{Capri:2014bsa}, a BRST-invariant action for the scalar field  $\phi^h$ is given by
\begin{equation}
S_{\sigma}=\sigma^{4}\,H_{\phi}(\phi^{h},A^{h})= g^{2}\sigma^4  \int d^{4}xd^{4}y\,  f^{abc}(\phi^h)^{b}(x)  [\mathcal{M}^{-1}(A^h)]^{ad}(x,y)  (\phi^h)^{e}(y)\,,  \label{nlsc}
\end{equation}
where the parameter  $\sigma^4$ plays a role analogous to that of the Gribov parameter $\gamma^4$ \footnote{Although we do not have a geometrical picture behind this parameter $\sigma^4$ akin to that of $\gamma^4$, i.e.~we do not have a horizon condition. Following footnote 7, it remains to work out a dynamical realization of $\sigma^4$.}. As shown in \cite{Capri:2014bsa}, the relevance of introducing the matter action \eqref{nlsc} lies in the interesting feature that, besides sharing a natural similarity with the horizon function $H(A)$ of the gauge field, it enables us to obtain a propagator for the scalar field which fits in a  nice  way the available lattice numerical data, see the discussion presented in \cite{Capri:2014bsa}.  Proceeding as before, the non-local term \eqref{nlsc} can be cast in local form
by introducing the auxiliary fields  ($\vartheta,\bar\vartheta$) and ($\theta,\bar\theta$)
\begin{equation}
 S_{\sigma}^{local} = \int d^4x\,\left\{-{\bar \vartheta}^{ac}  \mathcal{M}^{ab}(A^h) \vartheta^{bc}   + {\bar \theta}^{ac} \mathcal{M}^{ab}(A^h) \theta^{bc} + \sigma^2\,f^{abc} (\phi^h)^{a} (\vartheta^{bc} + {\bar \vartheta}^{bc}) \right\}\,.   \label{lsc}
\end{equation}
As done before, the action \eqref{lsc} is left invariant by the non-perturbative BRST transformations
\begin{eqnarray}
&& s_{\sigma^2} \vartheta^{ab} = \theta^{ab}\,,\qquad      s_{\sigma^2} {\bar \vartheta}^{ab} = 0\,, \nonumber\\
&&  s_{\sigma^{2}}\theta^{ab} = 0\,,\qquad s_{\sigma^2} {\bar \theta}^{ab} = {\bar \vartheta}^{ab} + \sigma^2 \,gf^{mnb}(\phi^h)^{m} \left(\frac{1}{\mathcal{M}(A^h)}\right)^{na}\,, \label{sssc}
\end{eqnarray}
with
\begin{equation}
s_{\sigma^2} S_{\sigma}^{local} =0 \;. \label{ss}
\end{equation}
Again, the transformations \eqref{sssc} can be localized in the same fashion as those of the pure gauge sector.

%%%
\subsection{Fermion fields in the fundamental representation}

The same construction can be applied to fermion fields in the fundamental representation. The BRST-invariant fermion field is now
\begin{equation}
 \psi^h = h^{\dagger} \psi\,.
\end{equation}
In this case, the invariant spinor action   is
\begin{equation}
S_{\varsigma}=\varsigma^{4}\,H_{\psi}(\psi^{h},A^{h})= -g^{2}\varsigma^4  \int d^{4}xd^{4}y\,  \bar(\psi^h)^{i}_{\alpha}(x) T^{a,ij}  [\mathcal{M}^{-1}(A^h)]^{ab}(x,y)  T^{b,jk}(\psi^h)^{k}_{\alpha}(y)\,,   \label{spi}
\end{equation}
where $i,j,k$ are the indices of the fundamental representation of SU($N$), $\alpha$ is the spinor index and  the parameter $\varsigma^4$ is analogous to $\sigma^{4}$ and  $\gamma^4$. As in the case of the scalar field, the term \eqref{spi} enables us to nicely reproduce the available lattice data on the quark propagator,  upon including as before a mass for the localizing auxiliary fermion fields \cite{Capri:2014bsa,Dudal:2013vha}.

%%%%%%%%%%%%%%%%%%

\section{Conclusion}\label{section:Finaldiscussion}

In the present work we have pursued the study of the nonperturbative nilpotent exact symmetry of the Gribov-Zwanziger formulation recently proposed in \cite{Capri:2015ixa}, in the Landau and linear covariant gauges.

The main issue which has been faced is that of the localization of both BRST operator and action derived in \cite{Capri:2015ixa}. As shown in Sections \eqref{loc-action} and \eqref{section:LocalBRST}, a complete local formulation can be achieved, as summarized by eqs.\eqref{gzlocalhorizon2},  \eqref{brstgammalocal}, \eqref{symm}. In addition of the auxiliary fields $({\bar \omega}, \omega, {\bar \varphi}, \varphi)$ already present in the original formulation, a second set of auxiliary fields $({\bar \zeta}, \zeta, {\bar \beta}, \beta)$, eq.\eqref{brstgammalocal}, as well as an auxiliary Stueckelberg field $\xi$, eq.\eqref{hxi}, have been introduced to obtain a fully local set up.

This novel formulation of the linear covariant gauges opens the possibility of new lines of investigation, as already partially reported in Sect.\eqref{section:Ward}. In particular, as a consequence of the Ward identities of the theory, the longitudinal component of the gauge field propagator has been proven to be completely determined to all orders, being given by $\frac{\alpha}{p^2}$, see eqs.\eqref{decgp},\eqref{G}. We emphasize that this result is in complete agreement with the recent lattice numerical simulations \cite{Cucchieri:2009kk,Bicudo:2015rma} as well as with the studies of the Dyson-Schwinger equations \cite{Aguilar:2015nqa,Huber:2015ria,Aguilar:2016vin}.

A second important consequence following from the  local Ward identities is the independence of the correlation functions of local BRST-invariant composite operators from the gauge parameter $\alpha$. For instance, this result implies that the masses of the glueballs already obtained in the Landau gauge \cite{Dudal:2010cd,Dudal:2013wja} remain the same when moving to the linear covariant gauges.

Concerning future investigations, we might quote the study of the infrared behaviour of the Faddeev-Popov ghost propagator in linear covariant gauges, the proof of the all order renormalizability of the new formulation, the analysis of the cohomology of the new local BRST operator, eq.\eqref{brstgammalocal}, and the possible identification of a suitable set of composite operators displaying the K\"all\'en-Lehmann spectral representation, a task of utmost importance for the understanding of the physical spectrum of the theory.  In addition, it will also be worthwhile to set up a dynamical framework to compute the various $d=2$ vacuum condensates needed to stabilize the Gribov-Zwanziger vacuum.

%%%%%%%%%%%%%%%%%%

\section*{Acknowledgements}

The Conselho Nacional de Desenvolvimento Cient\'{i}́fico e Tecnol\'{ó}gico (CNPq-Brazil), the Funda\c{c}\~ao de Amparo a Pesquisa do Rio de Janeiro (FAPERJ) \footnote{M.~S.~G.~is a level PQ-2 researcher under the program Produtividade em Pesquisa, 307905/2014-4 and is supported by the Jovem Cientista do Nosso Estado program - FAPERJ E-26/202.844/2015; L.~F.~P.~is supported by a BJT fellowship from the Brazilian program ``Ci\^{e}̂ncia sem Fronteiras'' (Grant No. 301111/2014-6).}, the Coordena\c{c}\~{a}̃o de Aperfei\c{c}̧oamento de Pessoal de N\'{i}́vel Superior (CAPES), the Pr\'{o}́-Reitoria de Pesquisa, P\'{o}́s-Gradua\c{c}\~{a}o e Inova\c{c}\~{a}o (PROPPI-UFF) and the Deutscher Akademischer Austauschdienst (DAAD) are acknowledged for financial support.

%%%%%%%%%%%%%%%%%%

\appendix

\section{Propagators of the elementary fields}\label{AppProps}

In order to calculate the tree-level two-point functions of the theory, let us first write the action
up to quadratic terms in the fields. For that matter, we consider the most general case of nonzero BRST-invariant
condensates $\langle  \left(\bar\omega^{ab}_{\mu}\omega^{ab}_{\mu} - \bar\varphi^{ab}_{\mu}\varphi^{ab}_{\mu} - \bar\beta^{ab}_{\mu}\beta^{ab}_{\mu}
+  \bar\zeta^{ab}_{\mu}\zeta^{ab}_{\mu}  \right) \rangle $ and $\langle (A^h)_{\mu}^a(A^h)_{\mu}^a \rangle $, as in the so-called refined Gribov-Zwanziger theory by adding
the term $S_{cond}$. An infrared cutoff for the Stueckelberg field $\xi$ is necessary, as
discussed in the Subsec.
\ref{subsection:StueckelbergMass}. This cutoff is added in a BRST-invariant way through the term $S_{IRR}$.
Let us start with the full action (\ref{gzlocalhorizon2}),
\begin{eqnarray}
S_{RGZ}^{LCG} &=& S_{YM} + S_{cond} + S_{IRR} + \int d^4x \left( \alpha\frac{b^ab^a}{2} + ib^a \partial_\mu A_\mu^a + {\bar c}^a (\partial_\mu D_\mu)^{ab} c^b \right) +\int d^4x \;\tau^{a}\,\partial_{\mu}(A^h)^{a}_{\mu}.
    \;\nonumber\\
&& \hspace{1cm} + \int\,d^4x \left(-\bar\varphi^{ac}_\mu{\cal M}(A^h)^{ab}\varphi^{bc}_\mu + \bar\omega^{ac}_\mu{\cal M}(A^h)^{ab}\omega^{bc}_\mu
+ g\frac{\gamma^2}{\sqrt{2}}\,f^{abc}(A^h)^a_\mu(\varphi_\mu^{bc} + \bar\varphi^{bc}_\mu)\right),
\nonumber\\
&& \hspace{1cm} +\int\,d^4x \left(-\bar\beta^{ac}_\mu{\cal M}(A^h)^{ab}\beta^{bc}_\mu + \bar\zeta^{ac}_\mu{\cal M}(A^h)^{ab}\zeta^{bc}_\mu
- g\frac{\gamma^2}{\sqrt{2}}\,f^{abc}(A^h)^a_\mu(\beta_\mu^{bc} + \bar\beta^{bc}_\mu)\right)  \;,
\label{gzlocalhorizon222}
\end{eqnarray}
supplemented by the Stueckelberg field mass regulator
\begin{eqnarray}
S_{IRR}&=&\int d^4x \frac{1}{2} s\left(\rho\xi^a\xi^a\right)
=\int d^4x \left( \frac{1}{2}M^4\xi^a\xi^a+\rho\xi^ac^a  \right)
\end{eqnarray}
and the dimension two condensates
\begin{eqnarray}
S_{cond}&=&\int d^4x\left\{\frac{m^2}{2}(A^h)_{\mu}^a(A^h)_{\mu}^a
+\mu^2\left(\bar\omega^{ab}_{\mu}\omega^{ab}_{\mu} - \bar\varphi^{ab}_{\mu}\varphi^{ab}_{\mu} - \bar\beta^{ab}_{\mu}\beta^{ab}_{\mu}
+\bar\zeta^{ab}_{\mu}\zeta^{ab}_{\mu}  \right)\right\}.
\end{eqnarray}

In order to extract the quadratic part of the action (\ref{gzlocalhorizon222}), we use the Stuckelberg field to write
the transversal gluonic field $A^h$ to first order in the fields as $(A^h)^a_{\mu}=A^a_{\mu}-\partial_{\mu}\xi^a$.
Besides, the Faddeev-Popov operator is trivially given by $\mathcal{M}^{ab}(A^h)=-\partial^2\delta^{ab}+{\cal O}(A^h)$.
The quadratic part of the action is then
\begin{eqnarray}
S_{\mathrm{RGZ}}^{quad} &=& S_{YM}^{(quad)}+
\int d^4x \left\{b^a\left(i\partial_{\mu}A^{a}_{\mu}+\frac{\alpha}{2}b^a\right)
+\bar{\varphi}^{ab}_{\mu}(\partial^2-\mu^2)\varphi^{ab}_{\mu}
+g\gamma^2f^{abc}(A^a_{\mu}-\partial_{\mu}\xi^a)\left(\varphi^{bc}_{\mu}+\bar{\varphi}^{bc}_{\mu}\right) +  {\bar c}^a \partial^2 c^a
\right.\nonumber\\
&&\left. - \bar{\omega}^{ab}_{\mu}(\partial^2-\mu^2)\omega^{ab}_{\mu} + \tau^a(\partial_{\mu}A^a_{\mu}-\partial^2\xi^a)+
\frac{1}{2}m^2\left(
A^a_{\mu}A^a_{\mu}-\xi^a\partial^2\xi^a-2A^a_{\mu}\partial_{\mu}\xi^a\right) +\frac{M^4}{2}\xi^a\xi^a +\rho\xi^ac^a
 \right.\nonumber\\
&&\left.\;
+\bar{\beta}^{ab}_{\mu}(\partial^2-\mu^2)\beta^{ab}_{\mu}
-g\frac{\gamma^2}{\sqrt2}f^{abc}(A^a_{\mu}-\partial_{\mu}\xi^a)\left(\beta^{bc}_{\mu}+\bar{\beta}^{bc}_{\mu}\right)
-  \bar{\zeta}^{ab}_{\mu}(\partial^2-\mu^2)\zeta^{ab}_{\mu}
\right\}.
\label{Squad}
\end{eqnarray}

Following the standard procedure, one finds the following set of tree-level propagators of the theory in momentum space
\begin{eqnarray}
\langle A^a_{\mu}(p)A^b_{\nu}(-p)\rangle &=& \frac{p^2+\mu^2}{p^4+(m^2+\mu^2)p^2+m^2\mu^2+2Ng^2\gamma^4}\delta^{ab}\mathcal{P}_{\mu\nu}
+\frac{\alpha}{p^4}p_{\mu}p_{\nu}
\\
\langle A_{\mu}^a(p)b^b(-p)\rangle&=& -\frac{p^2}{p^4+\alpha M^4}\delta^{ab}p_{\mu}
\\
\langle A_{\mu}^a(p)\varphi_{\nu}^{bc}(-p)\rangle=\langle A_{\mu}^a(p)\bar{\varphi}_{\nu}^{bc}(-p)\rangle
&=&\frac{g\gamma^2f^{abc}}{p^4+p^2(m^2+\mu^2)+m^2\mu^2+2Ng^2\gamma^4}\mathcal{P}_{\mu\nu}\\
\langle A_{\mu}^a(p)\xi^b(-p)\rangle&=& i\frac{\alpha\delta^{ab}}{p^4+\alpha M^4}p_{\mu}\\
\langle A_{\mu}^a(p)\tau^b(-p)\rangle&=&-i\frac{\alpha M^4}{p^2(p^4+\alpha M^4)}p_{\mu}\delta^{ab}
\end{eqnarray}
\begin{eqnarray}
\langle b^a(p)b^b(-p)\rangle &=&\frac{M^4}{p^4+\alpha M^4}\delta^{ab}
\\
\langle b^a(p)\xi^b(-p)\rangle&=&i\frac{p^2\delta^{ab}}{p^4+\alpha M^4}
\\
\langle b^a(p)\tau^b(-p)\rangle&=&-i\frac{M^4}{p^2}\delta^{ab}
\\
%%%%%%%%%%%%%%%%%%%%%%%%%%%%%%%%%%%%%%%%%%%%%%%%%%%%%%%%5
\langle\bar{c}^a(p)A^b_{\mu}(-p)\rangle &=&-i\frac{\rho\,\alpha}{p^2(p^4+\alpha M^4)}\delta^{ab}p_{\mu}\\
%%%%%%%%%%%%%%%%%%%%%%%%%%%%%%%%%%%%%%%%%%%%%%%%%%%%%%%%%%%%%%%%
\langle\bar{c}^a(p)b^b(-p)\rangle &=&i\frac{\rho}{p^4+\alpha M^4}\delta^{ab}\\
%%%%%%%%%%%%%%%%%%%%%%%%%%%%%%%%%%%%%%%%%%%%%%%%%%%%%%%%%%%%%%%%
\langle\bar{c}^a(p)\tau^b(-p)\rangle &=&\frac{\rho}{p^4+\alpha M^4}\delta^{ab}\\
\langle\bar{c}^a(p)\xi^b(-p)\rangle &=&\frac{\rho\,\alpha}{p^2(p^4+\alpha M^4)}\delta^{ab}\\
%\end{eqnarray}
%\begin{eqnarray}
\langle \bar\varphi_{\mu}^{ab}(p)\bar\varphi_{\nu}^{cd}(-p)\rangle=
\langle \varphi_{\mu}^{ab}(p)\varphi_{\nu}^{cd}(-p)\rangle
&=&\frac{g^2\gamma^4f^{abm}f^{mcd}}{(p^2+\mu^2)[p^4+p^2(m^2+\mu^2)+m^2\mu^2+2Ng^2\gamma^4]}\mathcal{P}_{\mu\nu}
\\
&&\hspace{-6cm}\langle \bar\varphi_{\mu}^{ab}(p)\varphi_{\nu}^{cd}(-p)\rangle
=\frac{g^2\gamma^4f^{abm}f^{mcd}}{(p^2+\mu^2)[p^4+p^2(m^2+\mu^2)+m^2\mu^2+2Ng^2\gamma^4]}\mathcal{P}_{\mu\nu}
-\frac{\delta^{ac}\delta^{bd}}{p^2+\mu^2}\delta_{\mu\nu}\\
\langle \varphi_{\mu}^a(p)\tau^b(-p)\rangle=\langle \bar{\varphi}_{\mu}^a(p)\tau^b(-p)\rangle
&=&-i\frac{g\gamma^2}{p^2(p^2+\mu^2)}p_{\mu}f^{abc}
\\
\langle\xi^a(p)\xi^b(-p)\rangle&=&\frac{\alpha\delta^{ab}}{p^4+\alpha M^4}
\\
\langle\xi^a(p)\tau^b(-p)\rangle &=&\frac{p^2}{p^4+\alpha M^4}\delta^{ab}
\\
\langle\tau^a(p)\tau^b(-p)\rangle&=&-\left\{\frac{m^2(p^4-\alpha M^4)+M^4p^2}{p^2(p^4+\alpha M^4)}
+\frac{2Ng^2\gamma^4}{p^2(p^2+\mu^2)}\right\}\delta^{ab}   \;,
%%%%%%%%%%%%%%%%%%%%%%%%%%%%%%%%%%%%%%%%%%%%%%%%%%%%%%%%%%%%%%%%%%%%%%%%%%
\end{eqnarray}
with $\mathcal{P}_{\mu\nu} = \left(\delta_{\mu\nu} - \frac{p_\mu p_\nu}{p^2} \right)$ being the transverse projector.
%%%%%%%%%%%%%%%%%%%%%%%%%%%%%%%%%%%%%%%%%%%%%%%%%%%%%%%%%%%%%%%%%%%%%

The theory also has a considerable set of vanishing propagators, given by all propagators involving
 either the $c$ or the the $\omega$ fields, except for
\begin{eqnarray}
\langle\bar{c}^a(p)c^b(-p)\rangle &=&\frac{1}{p^2}\delta^{ab}\\
\langle\bar{\omega}_{\mu}^{ab}(p)\omega_{\nu}^{cd}(-p)\rangle  &=&
-\frac{1}{p^2+\mu^2}\delta^{ac}\delta^{bd}\delta_{\mu\nu}.
\end{eqnarray}
%

%as well as
%
%\begin{eqnarray}
%&&\langle\xi^a(p)\varphi^{bc}_\mu(-p)\rangle=\langle\xi^a(p)\bar\varphi^{bc}_\mu(-p)\rangle=\langle b^a(p)\varphi^{bc}_\mu(-p)\rangle
%=\langle b^a(p)\bar\varphi^{bc}_\mu(-p)\rangle=\langle \bar{c}^a(p)\bar{c}^b(-p)\rangle 0.
%\end{eqnarray}
%%%%%%%%%%%%%%%%%%%%%%%%%%%%%%%%%%%%%%%%%%%%%%%%%%%%%%%%%%%%%%%%%%%%%%%%%%%%%%%%
%

The propagators involving the auxiliary fields $(\bar\beta,\beta)$ and $(\bar\zeta,\zeta)$ are
straightforwardly derived from those above by using the symmetry relations (\ref{discretesymm}). All
other propagators which have not been listed above are vanishing. Finally, let us recall that the parameters
$M$ and $\rho$ (which regularize the propagation of the Stueckelberg field) must be taken to zero at
the end of any actual calculation.

%%%%%%%%%%%%%%%%%%%%%%%%%%%%%%

\end{document}